\RequirePackage{amsmath,amsfonts}
\documentclass{article}
\usepackage{arxiv}

\usepackage{authblk}

\usepackage{graphicx}
\usepackage{subcaption}
\usepackage[abs]{overpic}
\usepackage{nohyperref}  
\usepackage{mathcomp}
\usepackage{caption}
\usepackage{units}
\usepackage{relsize}
\usepackage{centernot}
\usepackage{color}
\usepackage{transparent}
\usepackage{nicefrac}
\usepackage[strings]{underscore}
\usepackage{url}
\usepackage[titletoc]{appendix}

\newcommand{\comment}[1] { }

\newcommand{\vect}[1]{ \boldsymbol{\mathbf{#1}} }

\providecommand{\norm}[1]{\lVert#1\rVert}
\newcommand{\V}{ \mathcal{V} }
\newcommand{\W}{ \mathcal{W} }

\newcommand{\A}{ \mathcal{A} }

\newcommand{\eq}[1] {\begin{align} #1 \end{align}}
\newcommand{\eqlabel}[2] {
  \begin{align} \label{#1} #2 \end{align}
}
\newcommand{\Matrix}[1] { \begin{bmatrix} #1 \end{bmatrix} }

\begin{document}

\title{A momentum preserving frictional contact algorithm based on affine particle-in-cell grid transfers}

\author[a]{Michael Tupek}
\author[a]{Jacob Koester}
\author[a]{Matthew Mosby}
\affil[a]{Sandia National Laboratories}



\maketitle

\begin{abstract}
  An efficient and momentum conserving algorithm for enforcing contact between solid bodies is proposed.
  Previous advances in the material point method (MPM) led to a fast and simple, but potentially momentum violating, strategy for enforcing contact.
  This was achieved through a combination of velocity transfers between background and foreground grids, and a background grid velocity field update.
  We propose a modified strategy which ensures conservation of both linear and angular momentum with a novel use of the affine particle-in-cell (APIC) method.
  Two issues common to particle-in-cell based algorithms for contact are also addressed: material bodies tend to stick at a gap which is proportional to the grid spacing; and material points tend to stick together permanently when located within the same grid cell, making material rebound and friction challenging.
  We show that the use of APIC, combined with a grid transfer and momentum update algorithm results in contact being enforced at essentially zero gap.
  For the second issue, we propose a novel iterative scheme which allows particles interacting through the background grid to naturally separate after contact and enforce friction, while still satisfying momentum conservation.
\end{abstract}

\section{Introduction} \label{sec:Intro}

Simulating contact between solids remains as one of the more challenging aspects of finite deformation continuum mechanics simulations.
Here we propose a particle-in-cell based algorithm to enforce contact which is notable for its computational efficiency and simplicity.
The affine particle-in-cell (APIC) method is leveraged and applied in a novel way to efficiently enforce contact between two discretized solid bodies, while still allowing for material rebound and frictional sliding~\cite{Jiang2015,JIANG2017137}.
Global conservation of both linear and angular momentum is ensured at every step of the process.
Our approach is inspired by and related to a recently proposed contact algorithm for enforcing contact between finite element meshes using a material point method~\cite{Han_2019}.

The material point method (MPM) was first proposed in \cite{Sulsky:1994aa} and \cite{SULSKY1995236}, with the first strategy for contact in this setting introduced shortly thereafter in~\cite{BARDENHAGEN2000529}.
Various improvements to MPM contact can also be found in~\cite{nairn2005,doi:10.1002/nme.2981}.
Several authors have developed strategies to ensure conservation of linear and angular momentum in the MPM method, but focused on the partial differential equation discretization in the absence of contact~\cite{Love_2006,JIANG2017137,Fu_2017}.
For the method proposed here, MPM grid transfers are solely used to enforce contact, while the PDE discretization is based on traditional Lagrangian finite elements.
Linear tetrahedral elements are used, but the method works independent of the underlying weak form discretization.
A strategy for introducing discontinuities into the MPM was recently suggested in~\cite{MOUTSANIDIS2019584} and shows advantages for sliding and separating contact.
More broadly, material-point-like methods that use background B-spline grids are seeing increased use for a variety of mechanics, including shock, fracture, and nearly incompressible solids~\cite{Bazilevs:2017ab,Bazilevs:2017aa,Moutsanidis_2019}.

Looking beyond MPM methods, numerous strategies for enforcing contact have been proposed.
Of some relevance here is the work of \cite{PUSO20044891}, where the the source of angular momentum loss in Mortar contact methods is identified.
Mortar contact algorithms are generally held to be the most accurate for contact, but suffer from being somewhat challenging to implement, and potentially expensive to solve.
This difficulty is due to the need to solve for a Lagrange multiplier field at every time step during dynamic simulations.
Furthermore, standard Mortar contact implementations do not immediately conserve angular momentum during frictional sliding.
Another strategy for contact involves a series of increasing energy penalties which ensures no contact overlap, enforces linear and angular momentum conservation, and has good energy conservation properties~\cite{Harmon_2012}.
However, this approach can be computationally demanding and requires adapting the stable time step to resolve ever-increasing contact penalty frequencies.
In the context of meshfree methods, a strategy was recently presented in~\cite{Kamensky:2019aa} that ensures exact conservation of both linear and angular momentum, even when contact occurs with friction being enforced at a gap or with some material overlap.
A meshfree contact algorithm which uses a level-set reconstruction of the contacting surfaces that can result in smoother interactions was proposed in~\cite{Chi:2015aa}.
Another approach is based on G1 continuous surface patches to ensure smoother contact forces~\cite{doi:10.1002/nme.466}.

We also comment briefly on the primary motivation of the proposed algorithm for engineering application, namely computational efficiency.
Common strategies for contact involve a trade-off between accuracy, robustness, and computational efficiency.
The fidelity of contact ranges from the high accuracy of Mortar contact methods~\cite{PUSO20044891}, to, e.g., pinball contact algorithms~\cite{Belytschko:1991aa,Kamensky:2018aa} which use only penalty forces and fast sphere-sphere intersection searches.
The approach proposed here emphasizes computational efficiency in explicit dynamic simulations, while retaining accuracy where possible.

Efficiency is obtained here by completely avoiding the often expensive geometric search between nodes, faces, or elements, common to traditional contact strategies.
With these traditional contact searches, worst-case computational complexity scales super-linearly in the number of objects~\cite{10.5555/2383795.2383801}.
We replace this with two grid transfer operations which scale linearly in the number of nodes in contact.
Furthermore, the algorithm is trivially \emph{thread scalable}, which enables straightforward efficient implementation on many-core and graphics processing units (GPUs).
In the following, the focus is on the novel ideas which contribute to the proposed method, and its mechanical and numerical behavior.
Analysis of the computational advantages is left for further work.
Anecdotally the authors have observed significant performance advantages over classical search-based contact algorithms for explicit dynamic simulations in serial: about an order of magnitude.
These advantages become stronger on modern GPUs, as two orders of magnitude speedup or more is observed for sufficiently large problems.
 
In Section~\ref{sec:Formulation} we introduce the proposed contact formulation and motivate the various ingredients.
Semi-analytic and computational analyses of the proposed method can be found in Section~\ref{sec:Results}.
In addition, we numerically demonstrate energy convergence and the ability to accurately model sliding with frictional contact to verify the model in certain regimes.  
Finally, conclusions are in Section~\ref{sec:Conclusions}.

\section{Formulation} \label{sec:Formulation}

A novel formulation for automatically detecting and penalizing contact via background grid projections is proposed.  We build up the important pieces one at a time, starting with an explanation of what we call \emph{augury contact dynamics} for enforcing contact, and its time discretization.  This simplified starting point is then expaned upon to improve conservation and physical realism.  In particular, affine particle-in-cell grid transfers are utilized to conserve momentum and improve contact gap resolution, and a novel iteration procedure is introduced to allow for material rebound and friction to occur.

Before describing the proposed contact strategy, we introduce the general problem.  Consider any standard nodal finite element or meshfree discretizations of multiple solid mechanical bodies (see, e.g.~\cite{Belytschko:2000aa,Hughes:2012aa} for finite element approximations and, e.g., ~\cite{Nayroles:1992aa,Belytschko:1994aa,Sulsky:1994aa,Hillman:2014aa,Chen:2017aa} for meshfree).  For simplicity in this work we use standard linear triangular finite elements with one quadrature point per element.  Given a particular spatial discretization we identify all boundary nodes which may be involved in contact, and label these $P$ nodes with index $p$.  We will call these boundary nodes \emph{particles} to be consistent with PIC and MPM terminology.  The collection of \emph{all} nodes in the mesh will be indexed by $n$.  This nodal-based `foreground' discretization is where the discrete solution truly exists, and its time evolution is standard:
\eq{
  \dot{\vect{x}}_n &= \vect{v}_n \\
  \dot{\vect{v}}_n &= \vect{a}_n = \frac{\vect{f}_n}{{m}_n},
}
where $\vect{x}$ are the nodal coordinates, $\vect{v}$ the nodal velocities, $\vect{a}$ the nodal accelerations, $\vect{f}$ the external minus internal nodal forces, and $m$ are nodal masses.

\begin{figure}[tb]
  \centering
  \begin{overpic}[scale=0.5, unit=1mm]{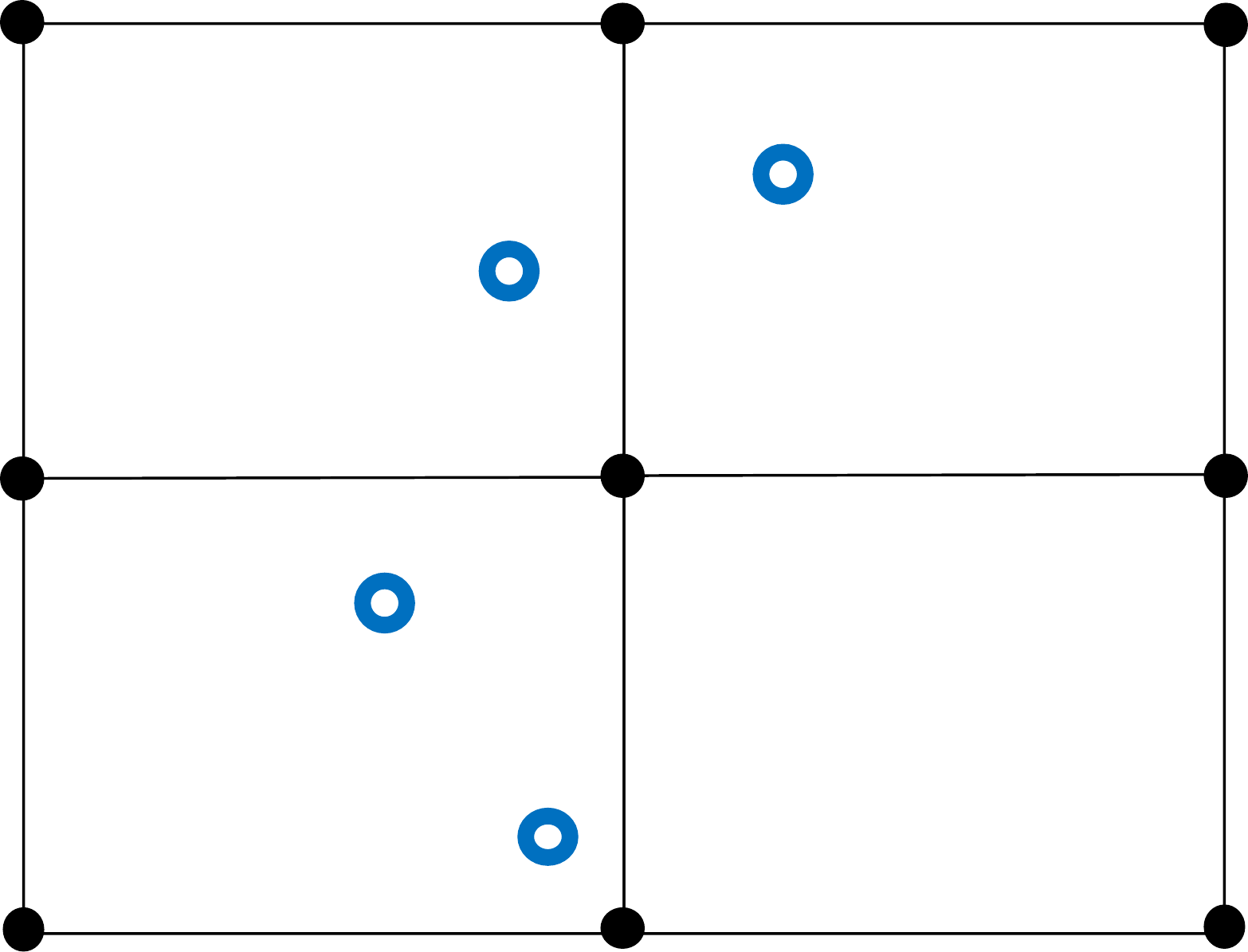}
    \put(41,27){
      \makebox[0pt]{$i$}
    }
    \put(32,39){
      \makebox[0pt]{$p$}
    }
  \end{overpic}
  \caption{Foreground particles and structured background grid.  The background grid nodes (solid black circles) are indexed by $i$, while the foreground (open blue circles) particles are indexed by $p$.}
  \label{fig:Grid}
\end{figure}

In addition to a standard nodal foreground discretization, we employ a structured Cartesian background grid.  Velocity and mass fields are approximated on this background grid using B-splines.  For simplicity, this work is restricted to using quadratic B-splines.  For details, see Appendix~\ref{subsec:spline}.  The use of background B-spline grids and other smooth background functions have become fairly common in MPM and PIC discretizations~\cite{Jiang2015,JIANG2017137,Bazilevs:2017ab,Bazilevs:2017aa,Moutsanidis_2019,Song_2020}.  In contrast to the approach taken here, typically the background grid in MPM hold the degree of freedom information, while the foreground mesh only tracks or advects with the background grid motion.  We label background grid nodes with index $i$.  A schematic depicting foreground particles and a background grid discretizations is shown in Figure~\ref{fig:Grid}.

\subsection{Augury dynamics}

In order to enforce contact in dynamical systems we propose a modification to the equations of motion which uses both a \emph{prediction} of where the system is heading and a model for the \emph{preferred} motion.  In particular, we derive what we call \emph{augury} acceleration correction terms which are used to take into account upcoming motions and push the system into a state which does not violate contact non-interpenetration constraints.
Given a system with position, velocity, and acceleration $\vect{x}_t$, $\vect{v}_t$, $\vect{a}_t$ at a given time $t$, a simple prediction of the velocity state at some time in the future $t+\tau$, $\tau>0$, is
\eq{
  \bar{\vect{v}}_{t+\tau} = \vect{v}_t + \vect{a}_t \tau,
}
where we use the notation subscript $t$ to denote the time dependence of these fields.
In addition, we provide a model which considers this predicted velocity and the current position state and produces a \emph{preferred} velocity field:
\eq{
  &\hat{\vect{v}}: \mathbb{R}^{3P} \times \mathbb{R}^{3P} \rightarrow \mathbb{R}^{3P} \\
  &\hat{\vect{v}}_{t+\tau} := \hat{\vect{v}}( \bar{\vect{v}}_{t+\tau}, \vect{x}_t ) 
}
where $P$ is again the number of boundary particles and 3 is the spatial dimension of the velocity vector.  This preferred future velocity function $\hat{\vect{v}}_{t+\tau}$ provides a model for the contact interaction behavior and different versions of it will be described in the upcoming sub-sections.
We modify the equations of motion by solving for the augury acceleration corrections $\hat{\vect{a}}_t$ which move the system toward this preferred velocity state:
\eq{
\hat{\vect{a}}_t = \frac{1}{\tau} \left( \hat{\vect{v}}_{t+\tau} - \bar{\vect{v}}_{t+\tau} \right),
}
which ensures
\eq{ \hat{\vect{v}}_{t+\tau} &= \bar{\vect{v}}_{t+\tau} + \hat{\vect{a}}_t \tau. \nonumber \\ &= \vect{v}_t + \left( \vect{a}_t + \hat{\vect{a}}_t \right) \tau }
We can simplify the equations slightly in the case where the preferred velocity is a linear function of the predicted velocity:
\eqlabel{eq:linear_preferred}{
  &\hat{\vect{v}}( \vect{v}_{t+\tau}, \vect{x}_t ) = \vect{H}_t \bar{\vect{v}}_{t+\tau},
}
where $\vect{H}_t = \vect{H}(\vect{x}_t)$ is just a matrix which depends only on the particle coordinates, and the augury acceleration correction is now
\eq{
  \hat{\vect{a}}_t = \frac{1}{\tau} \left( \vect{H}_t - \vect{I} \right)  \bar{\vect{v}}_{t+\tau}.
}
This modification to the equations of motion results in a system whose dynamics is altered gradually to track toward the desired state where the non-interpenetration conditions encoded in $\hat{\vect{v}}$ are approximately satisfied.

To motivate the use of these augury correction terms which use a prediction of the state at some point in the future, consider the limiting case of $\tau = 0^+$.  In this case we could choose to simply reset the system dynamics to the preferred state $\hat{\vect{v}}_{t^+}$.  However, this would require infinite contact forces to be applied and leads to unstable dynamics.  Allowing a finite value for $\tau$ overcomes this limitation.  In practice (see Section~\ref{sec:conservation}), we set $\tau$ proportional to the stable time step of the time discretized system.

\subsubsection{Augury dynamics using PIC grid transfers} \label{sec:augury_pic}

One option for defining the preferred velocity field $\hat{\vect{v}}_{t+\tau}$ is to perform a velocity field grid transfer from the foreground boundary particles to a background grid, followed by another transfer back from the background to the particles.\footnote{The particle-in-cell methods in \cite{Jiang2015,JIANG2017137,Han_2019} perform their time integration on the background grid.
As a result the particle-to-grid transfer occurs at a different time than the subsequent grid-to-particle transfer.}
The motivation for this is to exploit the natural and efficient contact detection properties of PIC-like grid transfers.  Two separate bodies with distinct particle velocity fields will project their velocity fields onto a common background grid where the velocity is necessarily continuous.  This has the side-effect of having the two bodies tending to stick together as they approach each other.  This is both good and bad, as contact is automatically enforced, but rebound cannot occur if the two bodies get sufficiently close.  For now we will focus on the positive feature and address the sticking limitation in Section~\ref{sec:material_separation}.

We review the standard particle-in-cell transfer employed here.
First, mass and momentum are transferred from particles to the background grid:
\eqlabel{eq:massToGrid}{ m_i = \sum_p N_i(\vect{x}_p) m_p , }
and
\eq{m_i \vect{v}_i = \sum_p N_i(\vect{x}_p) m_p   \vect{v}_p, }
where $\vect{v}_p$ are particle velocities, $m_p$ are particle masses, and the background grid spline shape functions $N_i$ for grid point $i$ are evaluated at particle locations $\vect{x}_p$.
The resulting grid velocities and grid masses are $\vect{v}_i$ and $m_i$.
Second, the velocity is interpolated back to the particles 
\eq{
  \tilde{\vect{v}}_{p} = \sum_i N_i(\vect{x}_{p}) {\vect{v}}_{i}.
}
This new velocity field on the particles is a smoothed version of the original field and the entire process can be thought of as a linear mapping
\eq{
  \tilde{\vect{v}} = \vect{H}^{PIC}(\vect{x}) \vect{v},
}
which means it satisfies the linear functional assumption in Equation~\eqref{eq:linear_preferred}.

The most basic augury dynamics strategy to enforce contact is to use this PIC smoothed field as the preferred velocity field:
\eq{
  \hat{\vect{v}}( \vect{v}_{t+\tau}, \vect{x}_t ) = \vect{H}^{PIC}(\vect{x}_t) \bar{\vect{v}}_{t+\tau}.
}
The corresponding acceleration correction is then
\eqlabel{eq:pic_accel}{
\hat{\vect{a}}_t = \frac{1}{\tau} \left( \vect{H}_t^{PIC} - \vect{I} \right) \left( \vect{v}_t + \vect{a}_t \tau \right).
}
The standard PIC grid transfers described here are known to conserve linear momentum~\cite{JIANG2017137}, so it follows that any velocity field $\vect{v}$ and its associated PIC smoothed field $\hat{\vect{v}} = \vect{H}^{PIC} \vect{v}$ have the same net linear momentum.
As a result the acceleration correction Equation~\eqref{eq:pic_accel} has no net effect on the total system linear momentum.  Consequently, we can think of the linear operator
\eq{
  \vect{G}^{PIC} := \vect{H}^{PIC} - \vect{I},
}
as one which maps any velocity field to a field with no net linear momentum.


\subsubsection{Augury dynamics using APIC grid transfers} \label{subsec:apic_formulation}

While the PIC transfers used in the previous section are known to conserve linear momentum, angular momentum conservation is not guaranteed.
Inspired by the affine particle-in-cell (APIC) approach proposed in~\cite{Jiang2015,JIANG2017137}, we consider APIC grid transfers, which are~\eqref{eq:massToGrid} together with
\eqlabel{eq:particleToGrid}{ 
m_i \vect{v}_i = \sum_p N_i(\vect{x}_p) m_p \left[
    \vect{v}_p + \vect{B}_p \vect{D}_p^{-1} (\vect{x}_i - \vect{x}_p) \right] ,
}
and
\eq{ 
\vect{D}_p
  = \sum_i N_i(\vect{x}_p) \left( \vect{x}_i - \vect{x}_p \right) \otimes \left(
  \vect{x}_i-\vect{x}_p\right),
}
where $\vect{B}_p$ is a tensor that depends on the velocity, defined shortly in equation~\eqref{eq:scaledVelGrad}.
Compared to more standard MPM or PIC transfers, the APIC approach enriches the particle velocities with additional degrees of freedom $\vect{B}_p$, which we will call the scaled velocity gradient since $\vect{B}_p \vect{D}_p^{-1}$ corresponds to a local approximation of the velocity gradient field.
The initial condition for $\vect{B}_p$ can be solved for from the initial conditions to ensure $\vect{B}_p \vect{D}^{-1}_p$ matches the initial velocity gradient.
The use of these extra degrees of freedom ensures that any affine motion in the particles will be exactly reproduced on the background spline grid.  
The grid-to-particle transfer for the APIC method is then
\eqlabel{eq:gridToParticle}{ 
\tilde{\vect{v}}_p = \sum_i N_i(\vect{x}_p) \vect{v}_i ,
}
and
\eqlabel{eq:scaledVelGrad}{ {\vect{B}}_p = \sum_i
  N_i(\vect{x}_p) \vect{v}_i \otimes \left( \vect{x}_i - \vect{x}_p \right).  
  }

When applied sequentially in this manner, these two transfer operations combine into a linear operator $\vect{H}^{APIC}: \mathbb{R}^{3P + 9P} \rightarrow \mathbb{R}^{3P + 9P}$, where $P$ is again the number of particles on the boundary, and the $3$ and $9$ are the number of components for the velocity vector and the scaled velocity gradient tensor.
We call $\vect{H}^{APIC}$ the APIC transformation operator.
Given a generalized velocity field
\eq{\V = \{ \vect{v}, \vect{B} \},
}
the linear operator $\vect{H}^{APIC}$ returns a new velocity field and scaled velocity gradient field which have been altered by the two grid transfers
\eq{
  \hat{\V} = \vect{H}^{APIC} \V.
  }
There are a few key properties of this linear operator.
The first is that affine velocity fields are exactly preserved under $\vect{H}^{APIC}$: 
\eq{ 
\check{\V} = \vect{H}^{APIC} \check{\V}, 
} 
for any $\check{\V}$ with
\eqlabel{eq:affine_v}{
  \vect{\check{v}}_p = \vect{v}_0 + \vect{L} \vect{x}, }
and
\eqlabel{eq:affine_b}{ {\vect{\check{B}}}_p = \vect{L}  \vect{D}_p, }
where $\vect{L}$ is a constant velocity gradient tensor.
A proof of this is provided in Appendix~\ref{subsec:linear_fields}.
In addition, linear and angular momentum are both preserved under the operation $\vect{H}^{APIC}$ for any $\V$.
This follows directly from the proofs in~\cite{JIANG2017137}, where both grid-to-particle and particle-to-grid transfers were separately shown to be both linear and angular momentum preserving.

Similar to the PIC case in Section~\ref{sec:augury_pic}, it follow from the above that we can define an operator
\eq{
  \vect{G}^{APIC}: \mathbb{R}^{3P + 9P} \rightarrow \mathbb{R}^{3P + 9P}, \nonumber
  \\ \vect{G}^{APIC} = \vect{H}^{APIC} - \vect{I},
}
which takes an arbitrary generalized velocity field and returns a generalized velocity field with no net linear or angular momentum.

\subsection{APIC augury dynamics for contact}

To enforce contact in explicit dynamic similations, a modification to the standard equations of motion for the spatially discretized system is presented.
This modification allows the use of the APIC transformation operator $\vect{H}^{APIC}$ to automatically detect and enforce contact.
Similar to the PIC case, $\vect{H}^{APIC}$ tends to return transformed velocity fields which discourage contact penetration due to the particles sticking together within cells, an effect which is common to many MPM methods.

For particles with current velocity $\vect{v}_t$, scaled velocity gradient $\vect{B}_t$, and mechanical acceleration $\vect{a}_t$ (the acceleration in the absence of contact), we can predict the generalized velocity at time $t+\tau$
\eq{
  \bar{\V}_{t+\tau} = \V_t + \tau \A_t,
}
where $\V$ is the generalized velocities as before, $\bar{\V}$ is the predicted generalized velocity in the future, and $\A$ is the corresponding generalized accelerations
\eq{
\A_p = \left\{ \vect{a}_p , \dot{\vect{B}}_p \right\}.
}
The mechanical scaled acceleration gradient rate is always set to zero, $\dot{\vect{B}}_p=\vect{0}$, as there is no classical internal work on that mode.  
APIC augury dynamics modifies the equations of motion by applying the APIC transformation to $\bar{\V}_{t+\tau}$ and solving for an additional acceleration term which supplies the APIC augury contact force correction
\eqlabel{eq:apic_accel}{
  \hat{\A}_t = \frac{1}{\tau} \left( \vect{H}^{APIC}-\vect{I}\right)  \bar{\V}_{t+\tau},
  }
  with $\vect{I}$ the identity operator.  This correction is the acceleration needed to reach the preferred generalized velocity, assuming all accelerations are constant over the time $t \rightarrow t+\tau$, and satisfies
\eq{
  \vect{H}^{APIC} \bar{\V}_{t+\tau} = \V_t + \tau \left( \A_t + \hat{\A}_t \right).
}
The resulting APIC augury equations of motion become
\eqlabel{eq:apic_Augury}{
\dot{\vect{x}}_t &= \vect{v}_t \nonumber \\
\dot{\V}_t &= {\A}_t + \hat{\A}_t,
}
where $\hat{\A}_t$ is by construction a linear and angular momentum free acceleration correction and $\A$ inherits its linear and angular momentum conserving properties from the foreground discretization.
Note that the operator $\vect{H}^{APIC}$ depends on the particle coordinates $\vect{x}_p$ and masses $m_p$ at time $t$.  As mentioned before $\dot{\vect{B}}_p = 0$, but $\vect{\dot{\hat{B}}}_p \neq 0$ in general. 
We integrate in time using the Newmark time stepping algorithm, with Newmark parameters chosen for explicit time integration and second order accuracy~\cite{doi:10.1002/zamm.19860660905}
\eqlabel{eq:Newmark}{
  {\V}_p^{n} &= {\V}_p^{n-\nicefrac{1}{2}} + \frac{1}{2} \Delta t \, {\A}_p^{n} \nonumber \\
  \vect{x}_p^{n+1} &= \vect{x}_p^{n} + \Delta t \, \vect{v}_p^{n+\nicefrac{1}{2}} \nonumber \\
  {\A}_p^{n+1} &= \left\{ {\vect{a}}(\vect{x}^{n+1})_p , \vect{0} \right\} + \hat{\A}_p^{n+1}  \nonumber \\
  {\V}_p^{n+\nicefrac{1}{2}} &= {\V}_p^{n} + \frac{1}{2} \Delta t \, {\A}_p^{n+1},
}
where $\vect{a}(\vect{x})_p$ is the mechanical acceleration model for particle $p$ arising from the foreground finite element discretization and, e.g., particular material modeling choices.  The augury correction $\hat{\vect{A}}_p^{n+1}$ is the particle acceleration correction arising from~\eqref{eq:apic_accel} at time step $n+1$.  Changing timesteps between steps can readily be included, but is omitted here for brevity.

\begin{figure*}[!htb]
  \centering
  \begin{subfigure}{\textwidth}
  \includegraphics[width=.95\textwidth]{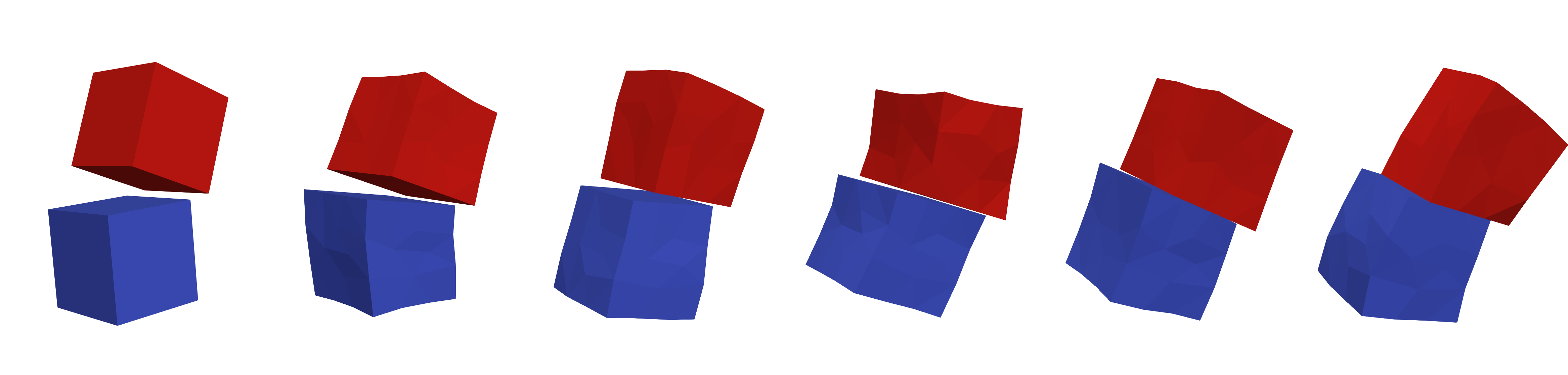}  
  \caption{Coarse mesh.}
  \end{subfigure}
    \begin{subfigure}{\textwidth}
  \includegraphics[width=.95\textwidth]{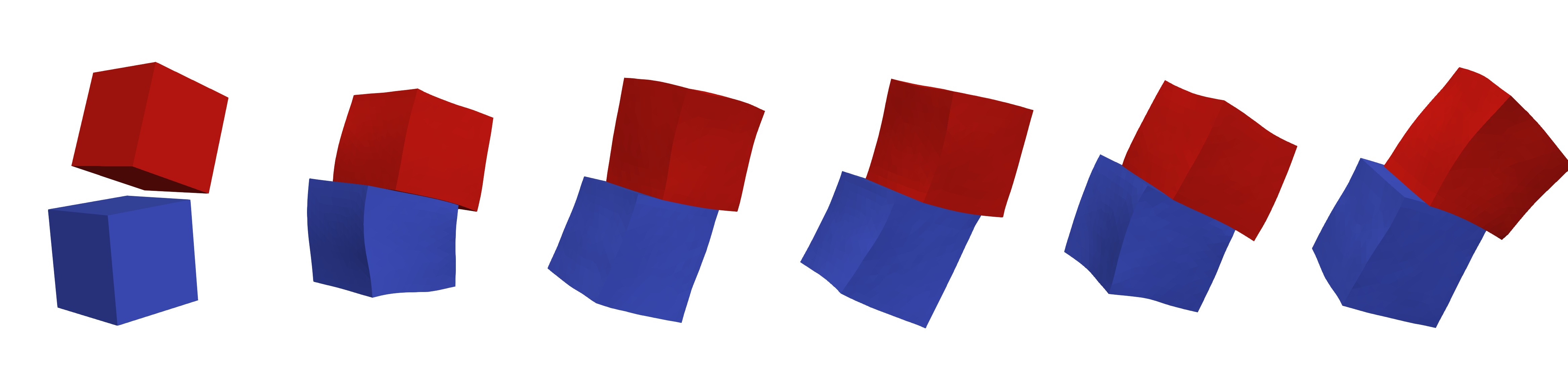}  
  \caption{Medium mesh.}
  \end{subfigure}
    \begin{subfigure}{\textwidth}
  \includegraphics[width=.95\textwidth]{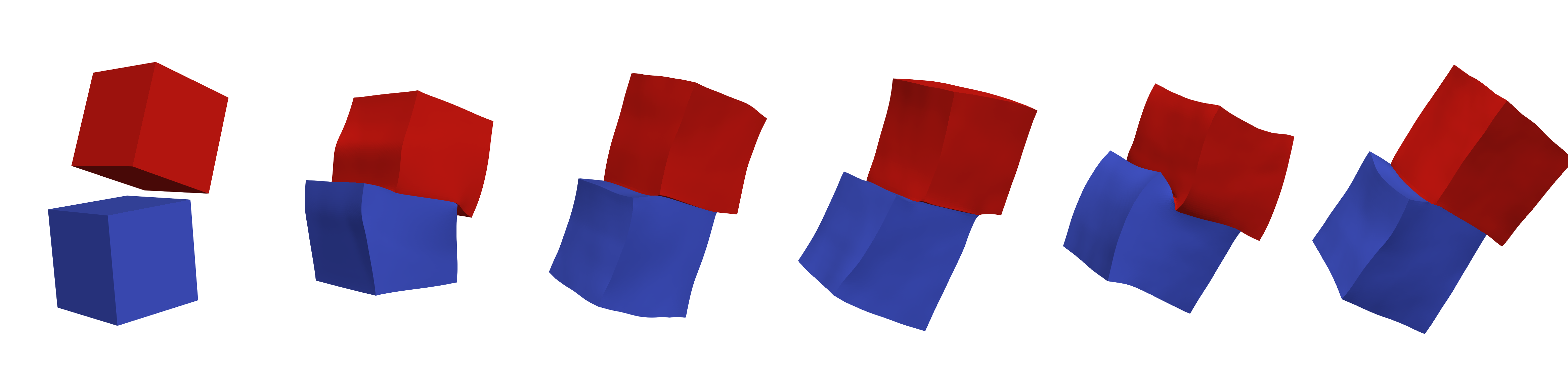}  
  \caption{Fine mesh.}
  \end{subfigure}
  \caption{Contacting blocks using APIC grid transfers.  The background grid size is set to approximately twice the element edge length.}
\label{fig:APIC_STICK}
\end{figure*}

\begin{figure}[!htb]
  \centering
  \begin{subfigure}{\textwidth}
  \includegraphics[width=.95\textwidth]{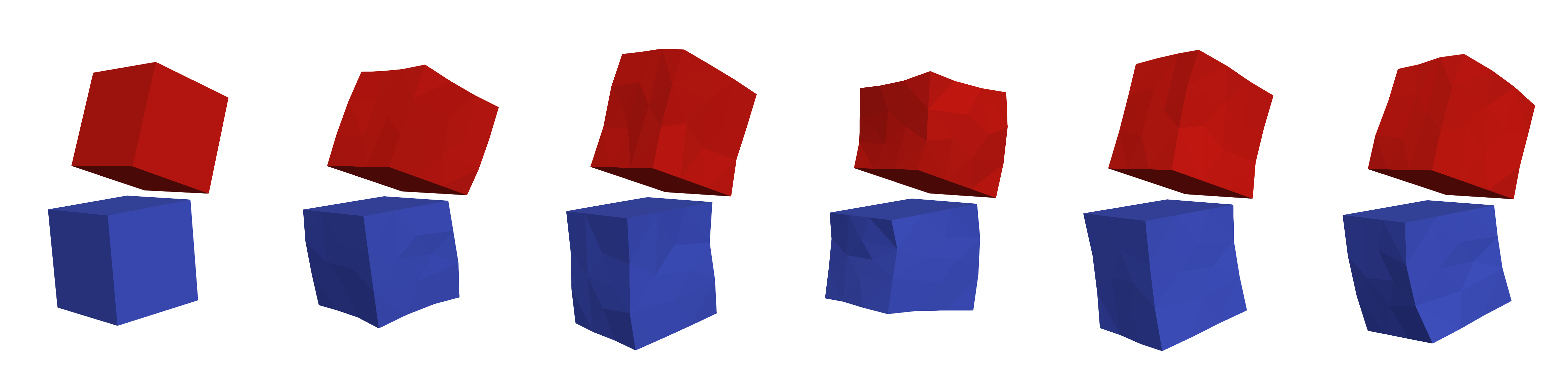}  
  \caption{Coarse mesh.}
  \end{subfigure}
    \begin{subfigure}{\textwidth}
  \includegraphics[width=.95\textwidth]{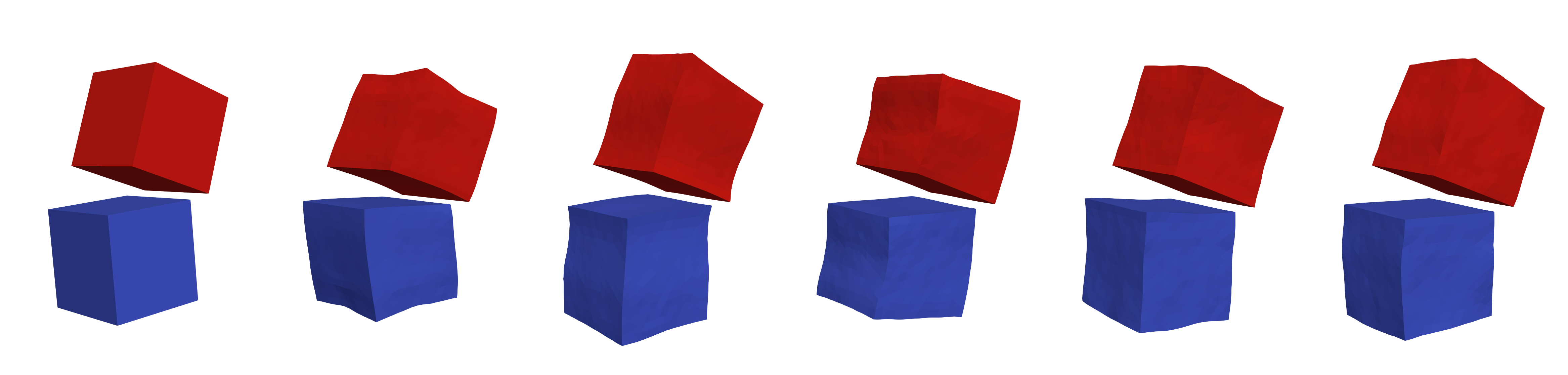}  
  \caption{Medium mesh.}
  \end{subfigure}
    \begin{subfigure}{\textwidth}
  \includegraphics[width=.95\textwidth]{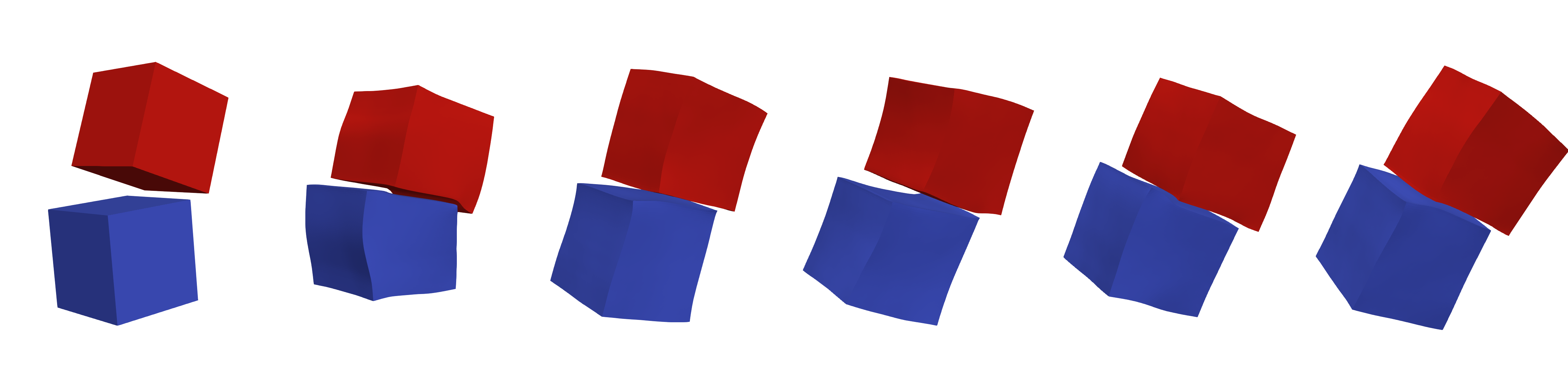}  
  \caption{Fine mesh.}
  \end{subfigure}
  \caption{Contacting blocks using standard MPM grid transfers.}
\label{fig:PIC_STICK}
\end{figure}

We call the resulting algorithm the \emph{APIC augury} algorithm for contact to distinguish it from the PIC algorithm for contact in~\cite{JIANG2017137} which uses a background grid update step to compute the momentum impulse applied to the foreground solids.  When applying this algorithm without the scaled velocity gradient term (e.g. by setting $\vect{D}_p= \vect{0}$), we call the resulting algorithm the \emph{PIC augury} algorithm.
Results using these formulations are shown in Figure~\ref{fig:APIC_STICK} where we simulate two blocks contacting at an angle.
Details of the simulation setup are in Section~\ref{sec:Results}.
We observe a few interesting features.
Upon contact, the two bodies become stuck together, as is expected for many PIC methods in which particles share a continuous background velocity field.
A less intuitive feature of these simulations is that the APIC augury algorithm results in the two bodies sticking at essentially zero visible gap/overlap.
This effect appears to be insensitive to both foreground mesh size and background grid size.
The reason for this behavior is not obvious, but it is desirable when contrasted with using a more classical PIC transfers (PIC augury dynamics), as shown in Figure~\ref{fig:PIC_STICK}.
For the lower order PIC grid transfers, the sticking occurs at a significant offset distance, proportional to the background grid size.  

In addition, we plot angular momentum over time for several cases in Figure~\ref{fig:AngularMomentumConservation}.
Only the APIC transfers guarantee angular momentum conservation for the simulation, as is expected from the construction of the algorithm.
Linear momentum is not shown, but it is conserved for both PIC or APIC augury dynamics.
This is in contrast to the method in~\cite{Han_2019} where even linear momentum conservation is not guaranteed in general.
It has also been shown that APIC transfers generally dissipate less energy than standard PIC transfers~\cite{Jiang2015,Fu_2017}, which is consistent with the results for this problem as shown and discussed further in Section~\ref{sec:Results}.

For the results presented, the background grid consists of uniform hexahedrons with edge lengths approximately twice the average finite element edge length.  If the background grid size is too small, it is possible for particles to pass each other without interacting through the background grid transfers.  This results in missed or inaccurate contact interactions.

\subsection{Material separation} \label{sec:material_separation}

The previous APIC correction is an improvement over traditional MPM transfers in terms of linear and angular momentum conservation, however there is still a clear limitation in that the two bodies completely stick after contact.
To address this we propose an iterative algorithm for calculating the augury acceleration correction which is based on a sequence of momentum-preserving grid transfer operations.
Consider a grid transfer linear operator $\vect{H}$ (either PIC or APIC depending on whether angular momentum conservation is considered important):
\eq{ \hat{\V} = \vect{H} \V }
which we can alternatively write as
\eq{ \hat{\V} = \V + \vect{G} \V, }
where we use $\vect{G} = \vect{H} - \vect{I}$.  The associated velocity correction associated with PIC grid transfers is 
\eq{{\Delta} \V = \hat{\V} - \V = \vect{G} \V. }
As $\hat{\V}$ and $\V$ have the same momentum (including angular if APIC is used), we conclude that $\Delta \V$ must have no net momentum, and this must hold for any $\V$.
As the range of $\vect{G}$ contains only zero net momentum fields, we can construct a new field with the same net momentum as $\V$ with any series of the form
\eq{
  \hat{\V} = \V + \vect{G}_0 \V + \sum_{n=1}^N \vect{G}_n \W_n, }
where $N$ is the number of so-called augury iterations, and $\W_n$ are arbitrary generalize velocity fields.
Once again the momentum of $\hat{\V}$ and $\V$ are equal provided that the $\vect{G}_n$ are all associated with momentum preserving grid transfers $\vect{H}_n$.  While we allow for the generality of potentially using different grid transfer operations for each term in this series, in practice we simplify to the case where each transfer operation is taken to be the same, $\vect{G}_i = \vect{G}$, $i=0,...,N$.

A strategy for allowing material rebound is to construct a series of this form with desirable mechanical properties.
In particular, we want to allow particle interactions through the background grid when bodies are coming into contact, but prevent their interaction when their velocities are opposing as in~\cite{Han_2019} to allow them to separate.
In the following, we compute a normal vector $\vect{n}_p$ for each particle $p$ in the current configuration by area weighting the outward normal vectors of the boundary faces adjacent to each finite element node/particle in contact.  

The iterative algorithm for enforcing frictionless sliding proceeds as follows.  At step $m$ with $m<N$ we compute the currently proposed modified velocity field for the particles as:
\eq{ \hat{\V}_m &:= \V + \vect{G} \V + \sum_{n=1}^{m} \vect{G}
  \W_n, \quad m < N. }
Consider $\Delta \vect{v}_{m,p} = \hat{\vect{v}}_{m,p} - \vect{v}_p$, for particle $p$, which is the proposed change in the particle velocity due to the PIC grid transfers up to this point.
For $\Delta  \vect{v}_{m,p} \cdot \vect{n}_p>0$, the proposed change to the velocity field acts to pull this particle away from its surface.
This is undesirable, as it indicates that particles are moving away from each other and we want the bodies capable of separation after contact.
To alleviate this, the undesirable component of the proposed velocity change can be iteratively smoothed out by setting
\eq{
  \W_{m+1,p} = \begin{cases} \Matrix{ \Delta \vect{v}_{m,p}  \\  \Delta \vect{B}_{m,p} } &\text{ if }  {\Delta \vect{v}}_{m,p}\cdot\vect{n}_p > 0 \\  \\ \Matrix{ \vect{0} \\ \vect{0} } &\text{ otherwise.} \end{cases}
}
For the next term in the series, $\W_{m+1}$ is acted upon by the operator $\vect{G}$.  This has the effect of gradually removing the contribution $\W_{m+1}$ from the series $\V_{m+1}$.  The intuition behind this effect is discussed below, but first a definition.

We define the \emph{quasi-null-space} of a grid transfer operator $\vect{H}$ as consisting of the generalized velocity fields $\V$ where $\hat{\V} = \vect{H} \V$ results in $\hat{\vect{v}} = \vect{0}$.  In other words, the generalized velocity fields which result in a zero physical velocity field (and any scaled velocity gradient) after the operation $\vect{H}$ has been applied.  For $\W$ in the quasi-null-space of $\vect{H}$, $\W^\prime = \vect{G} \W = \vect{H} \W - \W$ results in a field with
\eq{
  \vect{w}^\prime = -\vect{w},
}
so the operation $\vect{G}$ flips the sign of the physical part of the velocity for any field $\W$ in the quasi-null-space of the grid transfer linear operator $\vect{H}$.  A simple example of such a field is two equal mass particles with equal and opposite velocities interacting symmetrically in the same background cell.
In this case their momentum cancel, resulting in a transformed velocity field with zero particle velocities $\vect{v}_p = \vect{0}$, but may still have $\vect{B}_p \neq \vect{0}$.  Additionally, there are fields $\W$ in the quasi-null-space of $\vect{G}$, the most obvious example of this is any affine generalized velocity field, see equations~\eqref{eq:affine_v}--\eqref{eq:affine_b}.  The intuition is that the operator $\vect{G}$ tends to flip the sign of the physical part of the velocity field for fields which are non-affine, e.g., small wave-length features, but yields zero physical velocities for affine fields and near zero velocities for fields changing slowly over large wave-lengths.  

\begin{figure}[!htb]
  \centering
  \begin{subfigure}{\textwidth}
  \includegraphics[width=.95\textwidth]{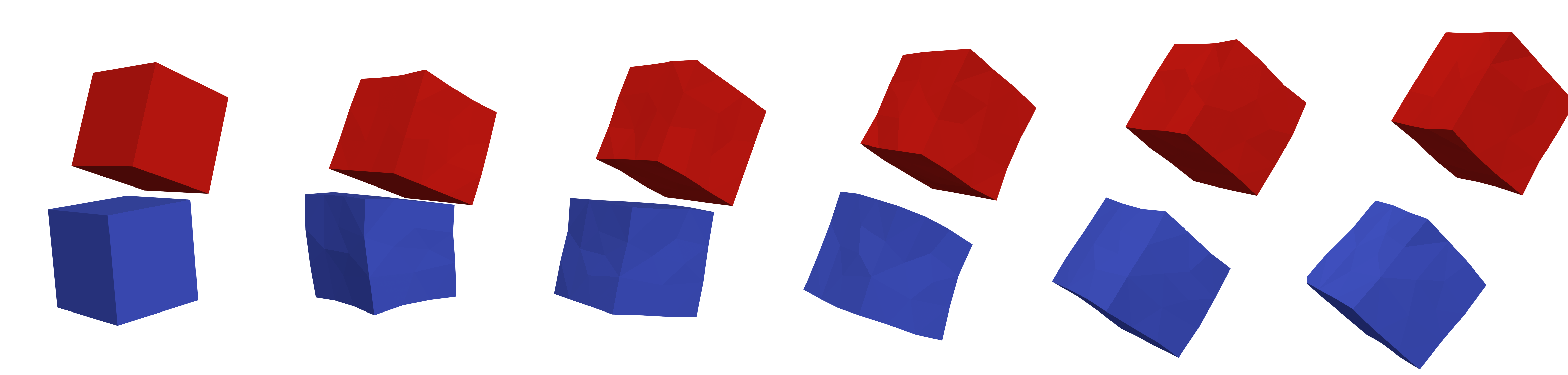}  
  \caption{Coarse mesh.}
  \end{subfigure}
    \begin{subfigure}{\textwidth}
  \includegraphics[width=.95\textwidth]{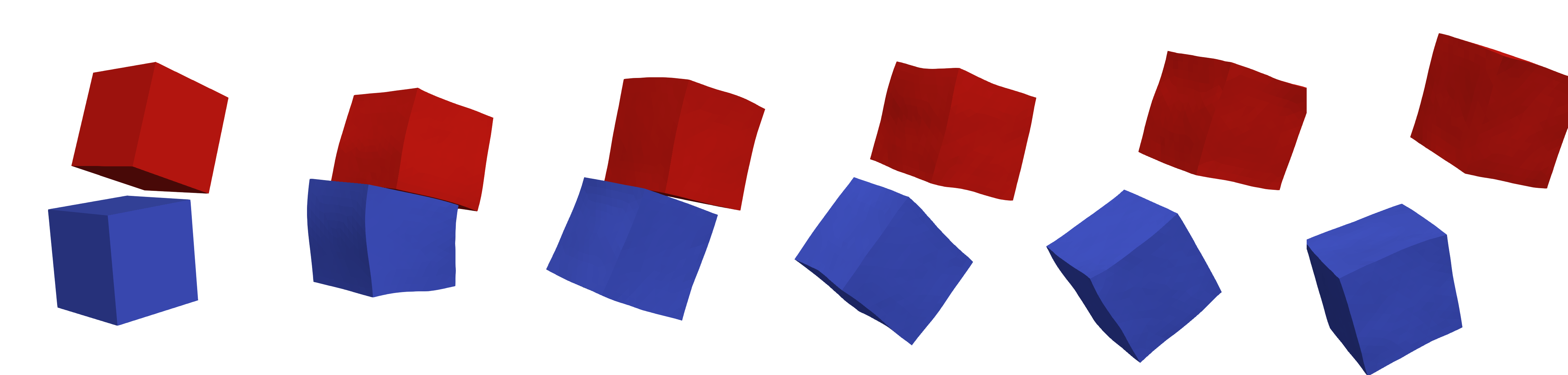}  
  \caption{Medium mesh.}
  \end{subfigure}
    \begin{subfigure}{\textwidth}
  \includegraphics[width=.95\textwidth]{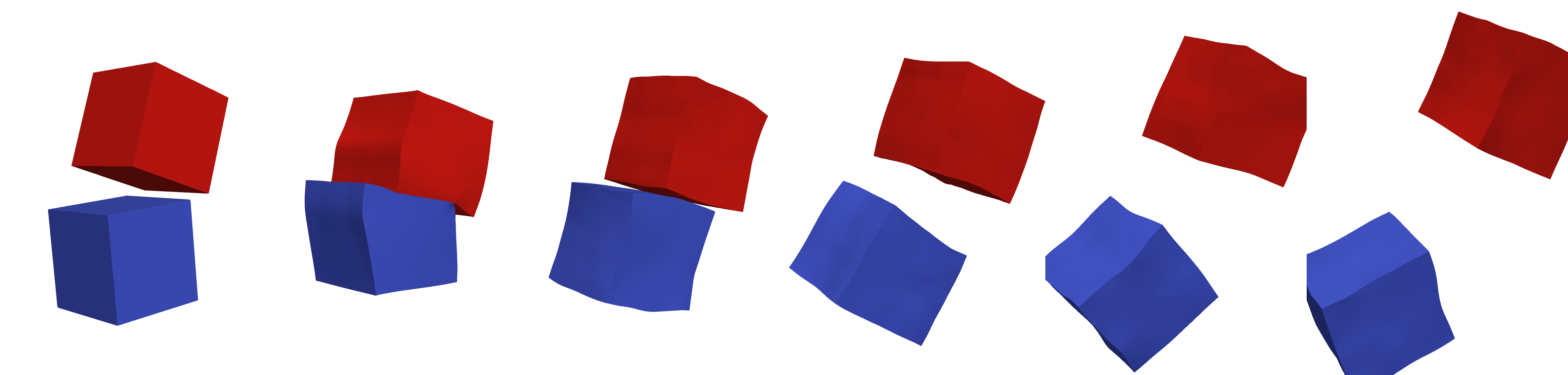}  
  \caption{Fine mesh.}
  \end{subfigure}
  \caption{Contacting blocks using APIC grid transfers and 1 iteration to enforce zero friction and material separation.}
\label{fig:APIC_BOUNCE}
\end{figure}

More analysis is needed to better understand the spectral properties of the operator $\vect{H}^{APIC}$ and how quickly we expect these iterations to converge in practice.
For this work, we only apply a single additional augury iteration, $N=1$.
We have not formally analyzed the convergence of the algorithm, but a small number of iterations produced reasonable results in the cases tried.
More work in this area is certainly warranted, but beyond the current scope.
Results repeating the earlier two block simulation using this iterative rebounding strategy are shown in Figure~\ref{fig:APIC_BOUNCE} using APIC transfers. We observe that the blocks bounce cleanly off each-other while sharply capturing the contact interaction, and that both linear and angular momentum are conserved.  Furthermore, while there is energy dissipation associated with the collision, the energy loss decreases under mesh and grid refinement.  This improved energy conservation behavior under refinement will be discussed further in Section~\ref{sec:conservation}.

\subsection{Sliding and friction}

We extend this iterative strategy to frictional contact, where the goal is to have particle velocity changes $\Delta \vect{\hat{v}}_{m,p}$ which satisfy
\eq{
  \Delta \vect{\hat{v}}_p \cdot \vect{n}_p &\leq 0 \\
 \norm{ \Delta \vect{\hat{v}}^t_{p} } &\leq \mu \Delta \vect{\hat{v}}_p \cdot \vect{n}_p,
}
where $\mu$ is the Coulumb friction coefficient, and $\Delta \vect{\hat{v}}^t_{p} = \Delta \vect{\hat{v}}_p \vect{n}_p \otimes \vect{n}_p$ is the velocity change perpendicular to the contact surface.
Here we drop the $m$ index to simplify syntax when needed.
At a given step $m$ we compute the undesirable portion of the generalized velocity, $\Delta{\V}^u_p$.  After evaluating
\eq{
  \Delta \vect{v}^t_p &= \Delta \vect{v}_p \vect{n}_p \otimes \vect{n}_p \\
  s^t_p &= \min(\norm{ \Delta \vect{v}^t_{p} }, -\mu \Delta \vect{v}_p \cdot \vect{n}_p) 
}
we compute the undesirable part of the physical velocity change
\eq{
  \Delta \vect{v}^u_p = \begin{cases} \Delta \vect{v}_p - \left( \Delta \vect{v}_p \cdot \vect{n}_p \right) \vect{n}_p - s^t_p \frac{\Delta \vect{v}^t_p}{ \norm{\Delta \vect{v}^t_p} }  &\text{ for } \Delta \vect{v}_p \cdot \vect{n}_p \leq 0 \\
 \Delta\vect{ v}_p  &\text{ otherwise } 
  \end{cases}
}
and the undesirable part of the scaled velocity gradient change
\eq{ \Delta \vect{B}^u_p = \Delta \vect{B}_p.}
We then set
\eq{ \W_{m+1,p} = \Matrix{ \Delta  \vect{v}^u_p \\ \Delta \vect{B}^u_p }. }
This iteratively enforces a rate form of the Coulombic friction model.
The choice to always treat the entire scaled velocity gradient difference as undesirable is debatable, but the intuition is to try and find a momentum conserving, contact enforcing solution with minimal changes to these enriched velocity gradient modes.

\section{Results} \label{sec:Results}

We apply this new contact algorithm on a few simple example problems to assess its behavior.
We begin by analyzing a two-particle system in 1D to gain insight into the algorithm's behavior.
Next, a block sliding with friction down an inclined plane is simulated to assess the accuracy and convergence properties of the friction model.
Finally, we demonstrate the smoothness properties of the proposed algorithm.

\subsection{Contact gap analysis for two particles in a background grid} \label{subsect:gap_analysis}

A well known behavior of MPM-like methods is that two bodies coming into contact tend to stick together at a finite distance proportional to the background grid size.
This is attributed to the continuous velocity field approximation in the background grid cells which results in the contacting particles following the same flow field.
Typically, when two particles enter the same cell the relative velocity between them will reach zero before the particles come into direct contact.
This error can artificially stiffen the response of multiple interacting bodies, and lead to larger total volumes than physically expected.
While this error does go away as the background grid is refined, it can remain as a significant source of error for certain applications.

\begin{figure}[tb]
  \centering
  \includegraphics[width=.7\textwidth]{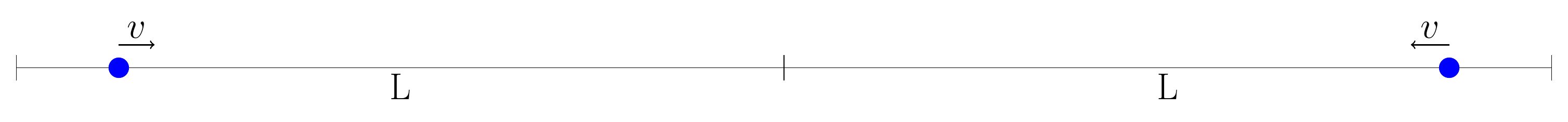}
  \caption{Two approaching particles interacting in a background grid of length $2L$.}
  \label{fig:two_particles}
\end{figure}

\begin{figure}[!htb]
\begin{subfigure}{.5\textwidth}
  \centering
  \includegraphics[width=.98\linewidth]{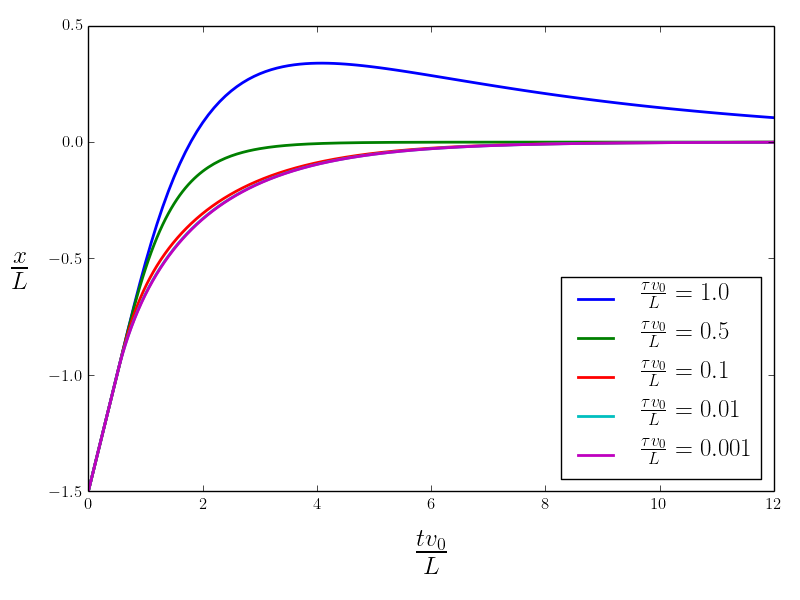}  
  \caption{Position vs time}
\end{subfigure}
\begin{subfigure}{.5\textwidth}
  \centering
  \includegraphics[width=.98\linewidth]{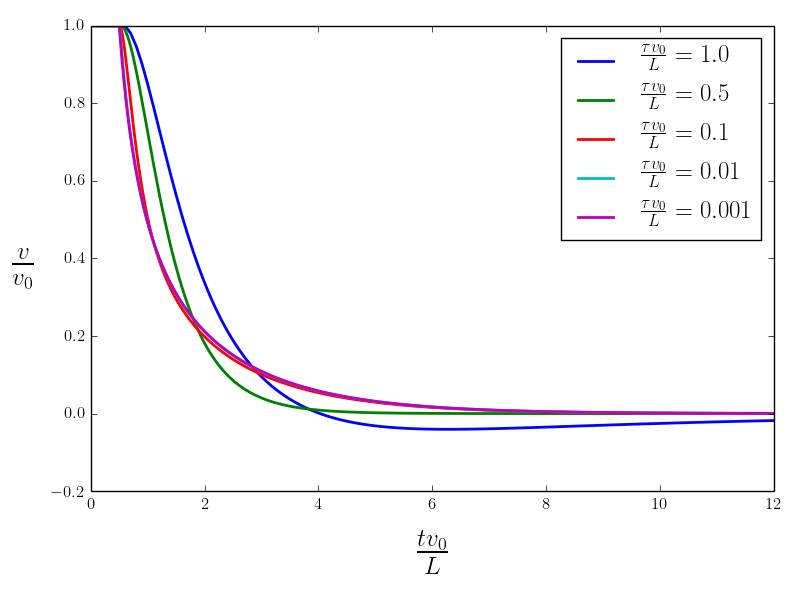}
  \caption{Velocity vs time}
\end{subfigure}
\\
\begin{subfigure}{.5\textwidth}
  \centering
  \includegraphics[width=.98\linewidth]{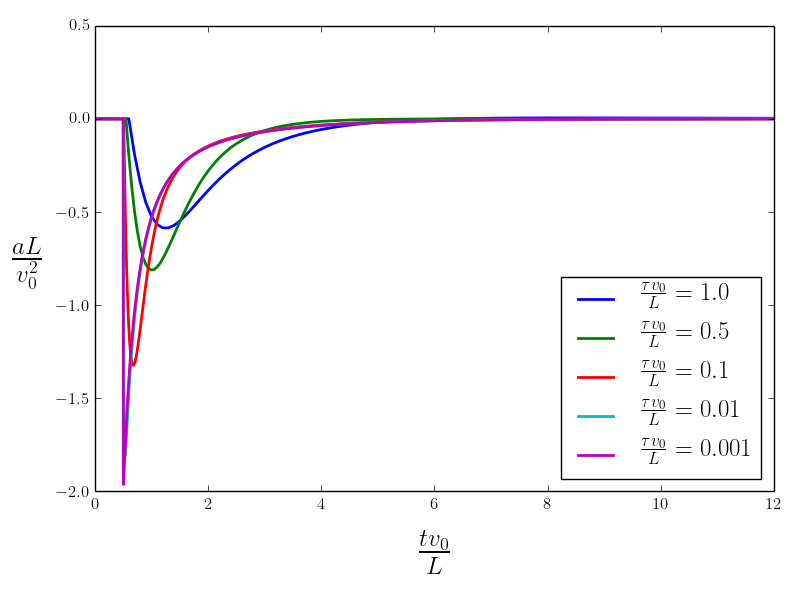}
  \caption{Acceleration vs time}
\end{subfigure}
\begin{subfigure}{.5\textwidth}
  \centering
  \includegraphics[width=.98\linewidth]{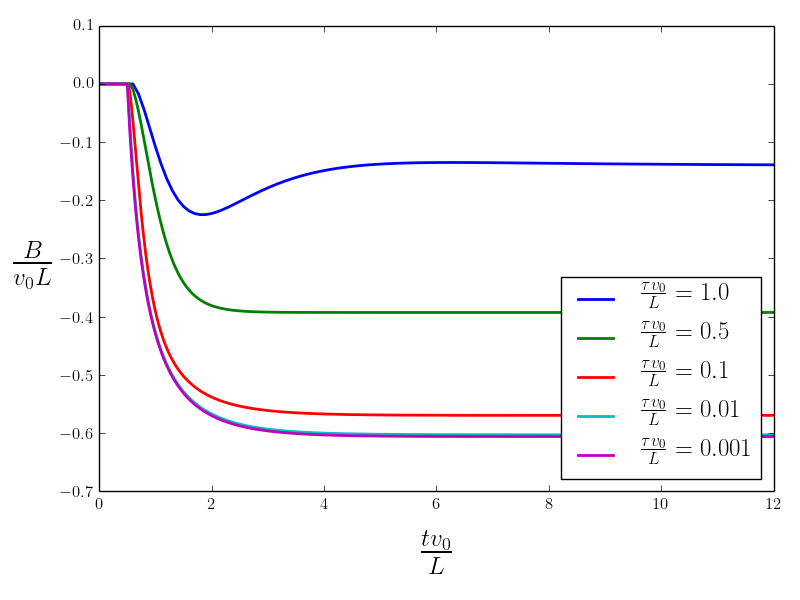}
  \caption{Scaled velocity gradient vs time}
\end{subfigure}
\caption{Particle dynamics of APIC augury equations with varying $\tau$.}
\label{fig:ApicAuguryDynamics}
\end{figure}

\begin{figure}[!htb]
\begin{subfigure}{.5\textwidth}
  \centering
  \includegraphics[width=.98\linewidth]{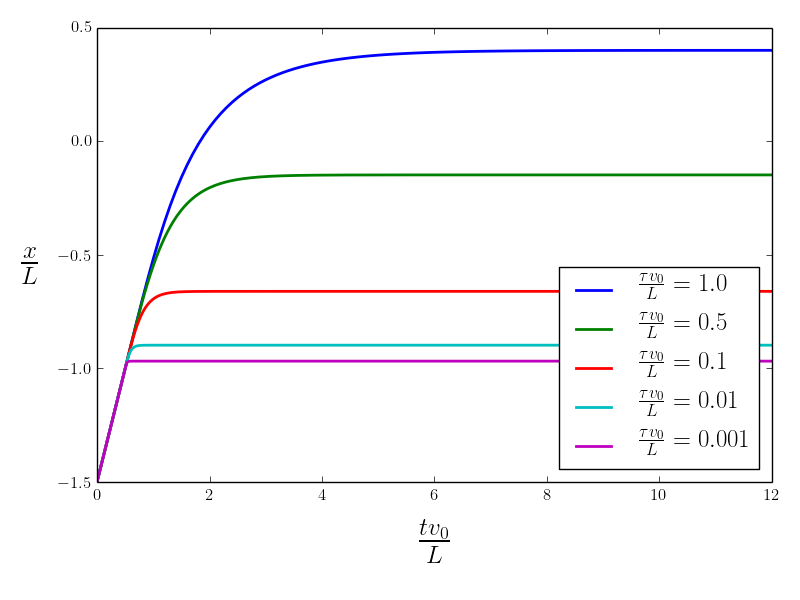}  
  \caption{Position vs time}
\end{subfigure}
\begin{subfigure}{.5\textwidth}
  \centering
  \includegraphics[width=.98\linewidth]{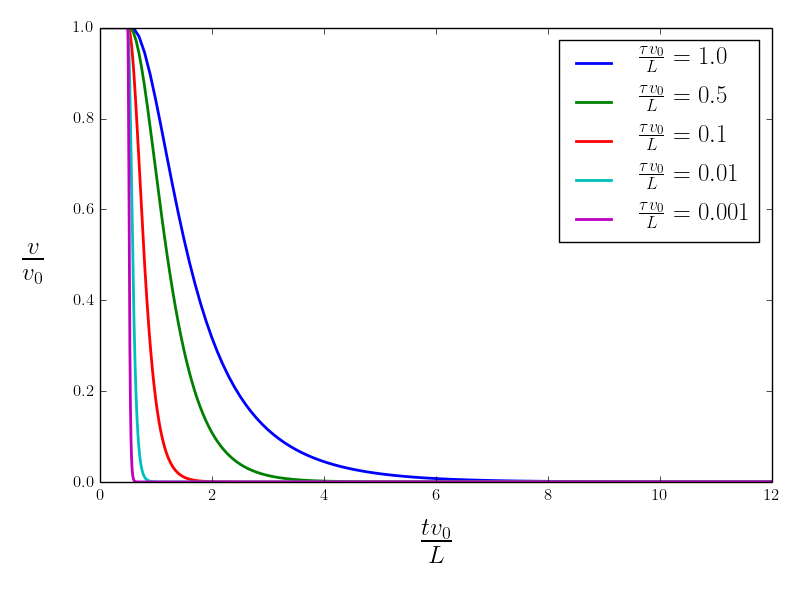}
  \caption{Velocity vs time}
\end{subfigure}
\caption{Particle velocity of PIC augury equations with varying $\tau$.}
\label{fig:PicAuguryDynamics}
\end{figure}

Remarkably, numerical experiments using the proposed APIC method found that this issue is significantly eliminated.
Contact (though sticky) automatically occurs at near zero visible gap.
This involved no additional algorithmic heuristics, and is an especially unexpected feature as the particles interacting through contact are completely unaware of each other's locations and velocities.
The only mechanism for communication between particles is through the background grid.
While we must admit that this feature is not yet fully understood, we provide a simple two particle analysis which helps unravel at least a piece of this fortunate mystery.
Consider the case where we have two particles of equal mass $m$ in 1D with initial coordinates $x_0$ and $-x_0$, and initial velocities $-v_{0}$ and $v_{0}$, respectively.
We consider the dynamics of these particle as they enter the same background cell which extends from $-L$ to $L$ (see Figure~\ref{fig:two_particles}).
When considering the APIC augury dynamics of particles in the absence of mechanical forces, the equations of motion~\eqref{eq:apic_Augury} simplify to
\eq{
  \dot{\vect{x}}_t &= \vect{v}_t \\
  \dot{\V}_t &= \frac{1}{\tau} \vect{G}^{APIC} \left(  \V_t  \right).
} 
We can non-dimensionalize by a change of variables
\eq{
  x &\leftarrow \frac{x}{L} \nonumber \\
  v &\leftarrow \frac{v}{v_0} \nonumber \\
  \tau &\leftarrow \frac{\tau v_0}{L} \nonumber \\
  B &\leftarrow \frac{B}{v_0 L} \nonumber
}
so that we only consider $v_{0} = 1$ and $L=1$.
To simplify the analysis we assume linear shape functions for the background grid approximation,\footnote{
  For the full-scale simulation later in the paper, we always employ quadratic B-Splines for the background grid approximation space.
  This 1D analysis is therefore only to gain intuition about a possible mechanism for the zero contact gap behavior.
}
$N_0(x) = \frac{1}{2}\left(1-x\right)$ and $N_1(x) = \frac{1}{2}\left(1+x\right)$.
By applying the grid transfers~\eqref{eq:particleToGrid} and~\eqref{eq:gridToParticle}, we find that APIC augury dynamics in this case simplifies to
\eq{
   \Matrix{
     \dot{x}_t \\
    \dot{v}_t \\
    \dot{B}_t
  } =
   \Matrix{
     1 & 0 \\
     x^2-1 & \;\; -x \\
     x^3-x & \;\; -x^2
  }
  \Matrix{
    v \\ B
  },
}
where $x$, $v$, $B$ are the non-dimensionalized coordinates, velocities, and scaled velocity gradient of the left particle.
Using standard PIC grid-transfers by setting $\vect{B}_p=\vect{0}$ in equation~\eqref{eq:particleToGrid} and removing its evolution, PIC augury dynamics is simply
\eq{
   \Matrix{
     \dot{x}_t \\
    \dot{v}_t \\
   } =
   \Matrix{
     v  \\
     v\left(x^2-1\right) \\
   }.
}
We gain intuition about the difference between the PIC and APIC augury dynamics by looking at the fixed points of the dynamical system.
With the PIC equations, we find if $v=0$, then $\dot{v} = 0$ for any $x$.
This indicates that the particles may come to rest at any distance from each-other.
For the APIC equations, things are a bit more interesting, as for any $B \neq 0$ if $v=0$, then $\dot{v} = 0$ only when $x=0$.
Furthermore, if $x=0$, then $\dot{B}=0$, so $v=x=0$, is the only fixed point of this dynamical system when $B\neq0$.
In other words, the particle can only come to rest once they are co-located, provided $B \neq 0$.
Numerical evidence suggests that this picture of the dynamics is qualitatively correct: standard PIC results in particles halting at a finite distance, while APIC results in particles halting only when they are co-located.  Some numerically computed solutions to these equations of motion are shown in Figures~\ref{fig:ApicAuguryDynamics} and Figures~\ref{fig:PicAuguryDynamics} for APIC and PIC dynamics respectively.  The key images are in Figures~\ref{fig:ApicAuguryDynamics}.a and Figures~\ref{fig:PicAuguryDynamics}.a, where the particle coordinates vs time are plotted for various values of the non-dimensional parameter $\nicefrac{\tau {v}_0}{{L}}$.
It is evident for PIC dynamics that the particles in general stick at a nonzero gap which is related to both the initial velocity and grid size.  For APIC, the particles come to rest only once they are both at $x=0$, and therefore the steady-state contact is enforced at zero gap.  We emphasize that this is a feature only of the steady-state behaviour.
It is still the case with APIC augury dynamics that the particles begin to have forces applied between them as soon as they enter the same grid cell and well before then finally stop at $x=0$.  The remainder of images in Figures~\ref{fig:ApicAuguryDynamics} and Figures~\ref{fig:PicAuguryDynamics} show the solution for the other terms in the equations of motion for completeness. 

These results indicate that the essential ingredient for obtaining perfect contact gaps with background cell methods is the addition of the APIC scaled velocity gradient term, while the background grid shape functions may be of secondary importance.
In fact, the dynamics for an arbitrary background grid approximations space are generically written: 
\eq{
  \Matrix{
     \dot{x}_t \\
    \dot{v}_t \\
    \dot{B}_t
  } =
   \Matrix{
     1 & 0 \\
     G_{11}(x) & G_{12}(x) \\
     G_{21}(x) & G_{22}(x)
  }
  \Matrix{
    v \\ B
  }.
}
So long as $G_{21} \neq 0$ for some $x$ is the path of the dynamics (to induce a change in $B$), and $G_{12}(x) = 0$ if and only if $x=0$ and $G_{22}(0) = 0$, the same arguments as above will hold.  In higher dimension, we believe smoothness of the background grid also likely plays an important role in the robustness and quality of the solution.

The analysis in this section has a few important caveats that are worth mentioning.
The first is that as the initial velocity dimensionless parameter $\nicefrac{\tau {v}_0}{{L}}$ increases, the interacting particles can actually pass each other, lose contact, and go unstable.
This highlights a possible limitation for the algorithm for high velocities relative to grid size.
It is therefore essential in practice to choose a $\tau$ which would ensure the system dynamics remains stable.
 Another caveat is that it is not straightforward to reproduce a similar dynamical system behavior using a background 1D B-spline grid (namely sticking at zero gap), so the conclusions drawn from this analysis may be limited.
 The mechanism for zero-gap sticking presented here may indeed be a large part of the explanation, but an understanding of the full mechanism in 3D using background smooth spline grids requires additional investigation.  We speculate that there is also a multi-dimensional aspect of this phenomena which is essential to explain the zero-gap sticking behavior in the 3D simulations.   Despite these limitation, the general trend of sticking at nearly zero gap, and loss of contact enforcement at high velocity to mesh-size ratios is consistent with 3D simulation results.

\subsection{Evaluation of conservation of momentum and energy} \label{sec:conservation}

To assess conservation of angular momentum and energy in a reasonably general scenario, we simulate two blocks impacting each other at an angle.  Both blocks are 3x3x3 cubes.  Block 1 is axis aligned and centered around the origin.  Block 2 begins coincident with Block 1, is first translated by 3.8 in y, then rotated by 22$^\circ$ about the global x-axis. The material property of both blocks is a density of 2700, a Young's modulus of $1 \times 10^8$, and a Poisson's ratio of 0.3.  A compressible Neo-Hookean hyperelastic material model is used.  Block 1 has an initial velocity of $v_0 = (0,35,0)$, while block 2 has an initial velocity of $v_0 = (-15, -35, -15)$.
Different foreground mesh sizes are simulated with $h=2^k$ for $k=\{0,-1,-2,-3,-4\}$.  For a given $h$, the background grid size is always set to $2h$.   Similarly the fixed time step is taken as $\Delta t = 1 \times 10^{-4} h$, and the augury time for each case is taken as $\tau = \Delta t$.  Example shap-shots of these simulations are shown in Figure~\ref{fig:APIC_BOUNCE}.

\begin{figure}[!htb]
\begin{subfigure}{.5\textwidth}
  \centering
  \includegraphics[width=.98\linewidth]{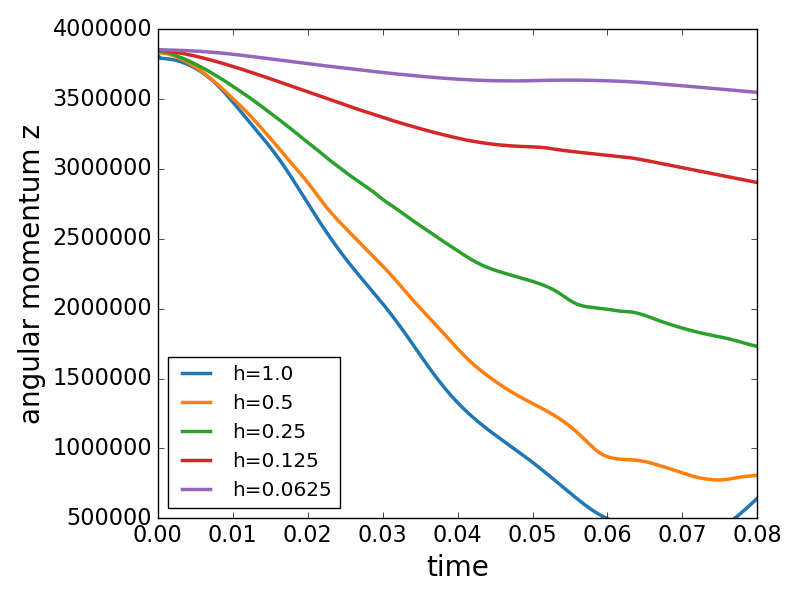}  
  \caption{PIC with no augury iteration.}
 \end{subfigure}
\begin{subfigure}{.5\textwidth}
  \centering
  \includegraphics[width=.98\linewidth]{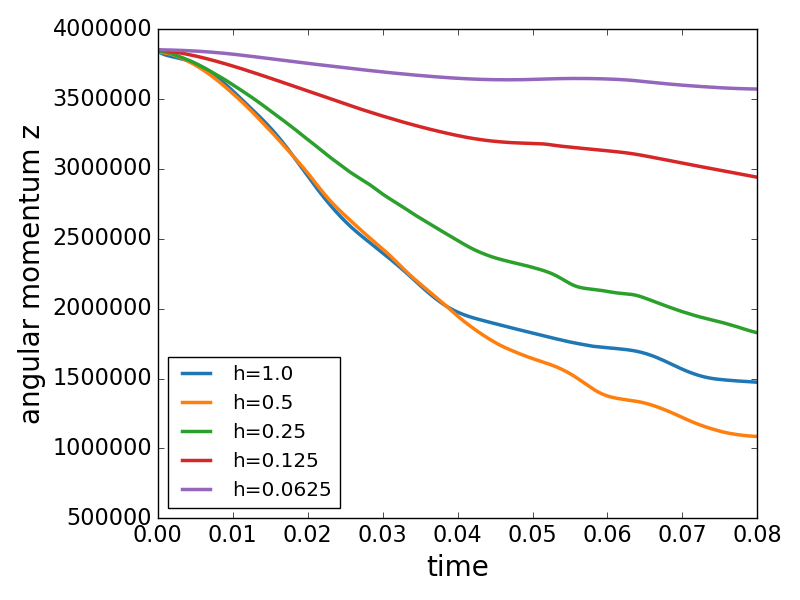}
  \caption{PIC with one augury iteration.}
 \end{subfigure} \\
\begin{subfigure}{.5\textwidth}
  \centering
  \includegraphics[width=.98\linewidth]{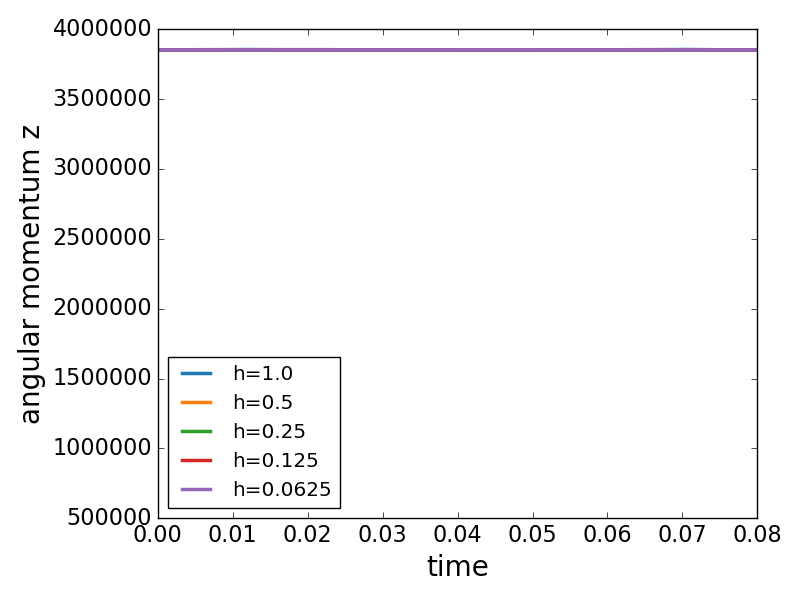}  
  \caption{APIC with no augury iteration.}
\end{subfigure}
\begin{subfigure}{.5\textwidth}
  \centering
  \includegraphics[width=.98\linewidth]{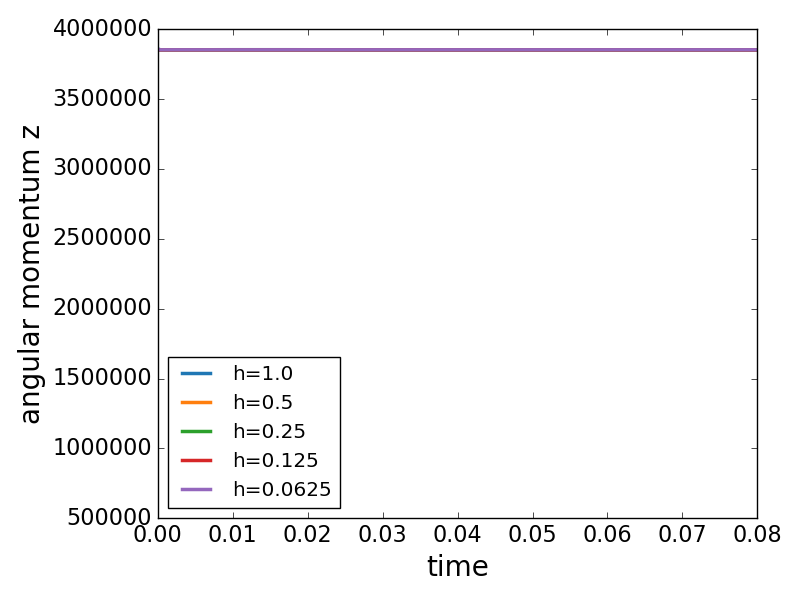}
  \caption{APIC with one augury iteration.}
\end{subfigure} 
\caption{Angular momentum conservation with different grid-transfer strategies.}
\label{fig:AngularMomentumConservation}
\end{figure}

\begin{figure}[!htb]
\begin{subfigure}{.5\textwidth}
  \centering
  \includegraphics[width=.98\linewidth]{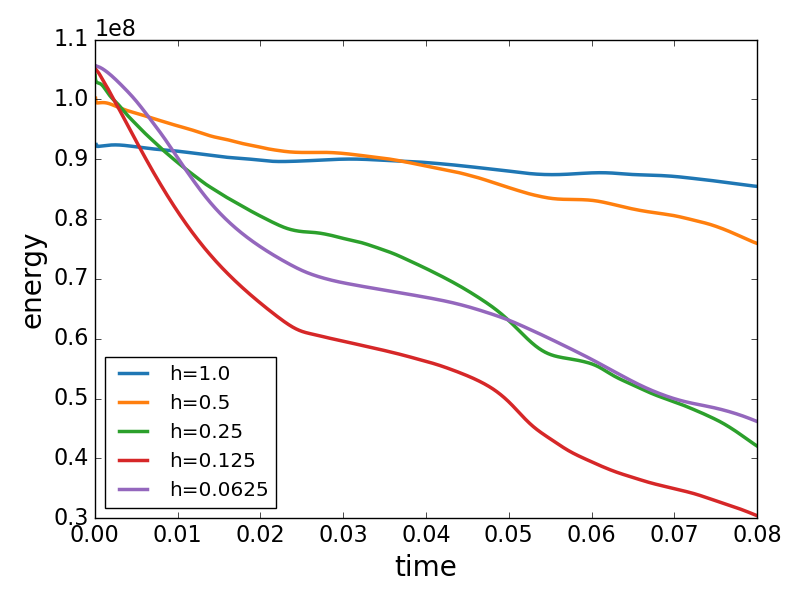}  
  \caption{PIC with no augury iteration.}
\end{subfigure}
\begin{subfigure}{.5\textwidth}
  \centering
  \includegraphics[width=.98\linewidth]{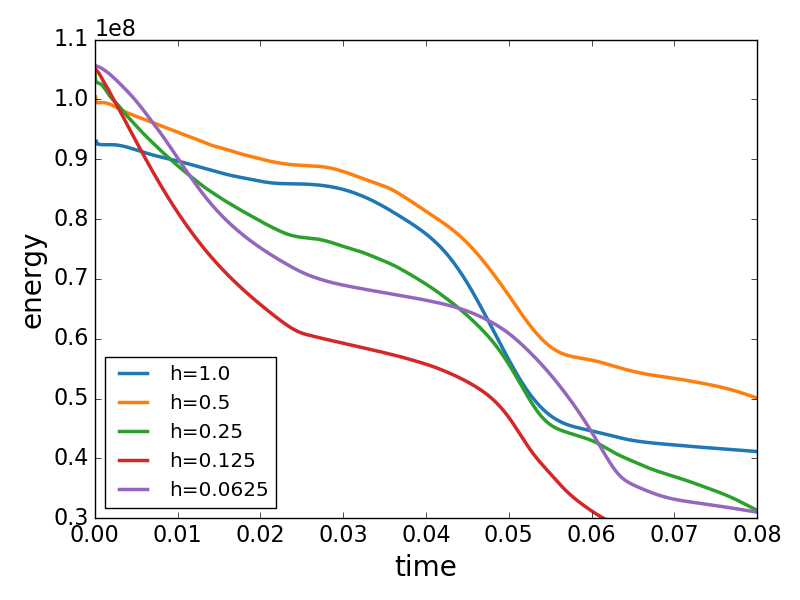}
  \caption{PIC with one augury iteration.}
\end{subfigure} \\
\begin{subfigure}{.5\textwidth}
  \centering
  \includegraphics[width=.98\linewidth]{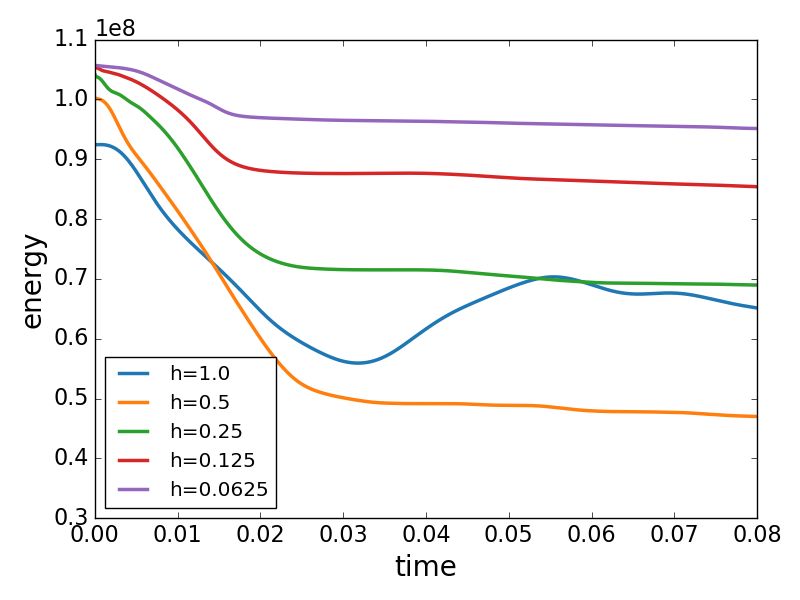}  
  \caption{APIC with no augury iteration.}
\end{subfigure}
\begin{subfigure}{.5\textwidth}
  \centering
  \includegraphics[width=.98\linewidth]{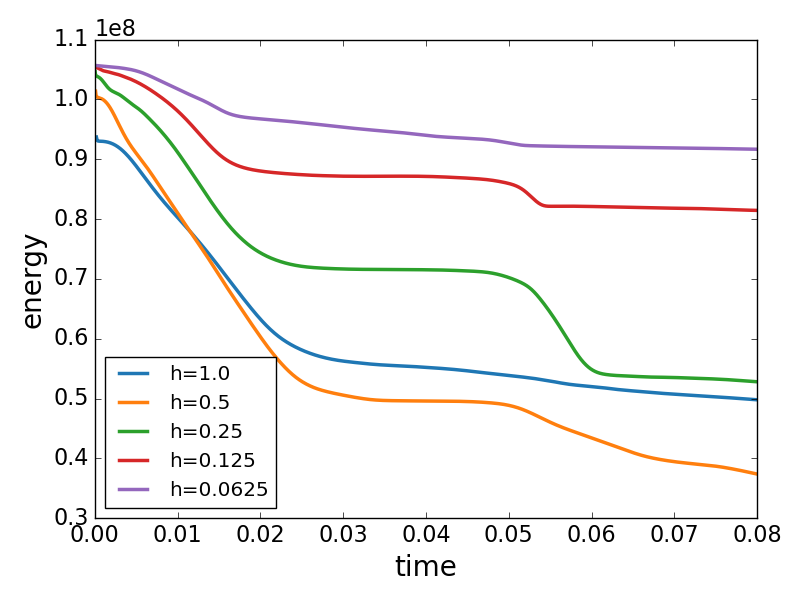}
  \caption{APIC with one augury iteration.}
\end{subfigure} 
\caption{Total energy conservation with different grid-transfer strategies.}
\label{fig:EnergyConservation}
\end{figure}

Results showing conservation of angular momentum for different formulations are shown in Figure~\ref{fig:AngularMomentumConservation}.  Only the APIC versions exactly conserve angular momentum.  
Results for total energy conservation are shown in Figure~\ref{fig:EnergyConservation}.  
We see that as the mesh is refined, the APIC versions do a significantly better job of conserving energy, though all are somewhat dissipative.\footnote{For computing kinetic energy, we are presently neglecting a contribution associated with the scaled velocity gradient field on the finite element mesh.  This means that the energy conservation for the APIC cases is actually a bit better than indicated, so this omission does not diminish the essential conclusions presented here.  In addition, this contribution to the total kinetic energy is expected to vanish as the mesh is refined.}  
For APIC, the dissipation is due to the fact that any component of the velocity field which is non-affine will be damped slightly due to the APIC grid transfers.  For PIC, any non-constant velocity fields will be damped over time.   The difference in initial energies is due to the simulation setup where the two bodies are instantly interacting through grid transfers, so their initial energies are slightly damped.  This effect decreases as the mesh and background grids are refined.

While the proposed method shows good promise for the regimes demonstrated here, we acknowledge that there are regimes in which it is still unstable.  This typically manifests as the contacting objects eventually losing their repulsive contact forces, interpenetrating and falling into one another.  This is currently a limiting factor in the practical use of this MPM-based contact algorithm for certain applications.  This means additional efforts are required to fully exploit the efficiency of this style of contact, and we believe the analysis in~\ref{subsect:gap_analysis} may help shed light on this current limitation.  Another possibility is that the iterative strategy for enforcing friction and rebound may be too aggressive at adjusting nodal velocities and may incidentally be removing desirable contact forces.  Despite these remaining challenges, the novel ideas and approaches presented here provide a unique and promising step forward relative to previous MPM contact strategies.

\clearpage

\subsection{Frictional sliding down a ramp}

\begin{figure}[!htb]
\centering
\includegraphics[width=.98\linewidth]{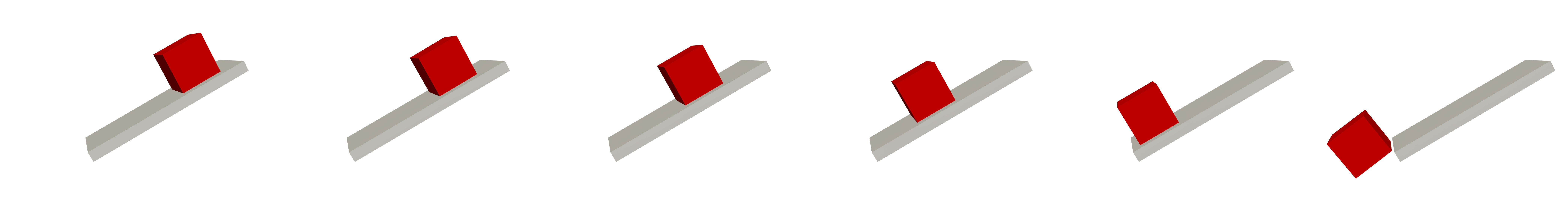}  
\caption{Snapshots of block sliding down a ramp using APIC and $\mu=0.2$.}
\label{fig:SlidingBlockSnap}
\end{figure}

\begin{figure}[!htb]
  \centering
\begin{subfigure}{.6\textwidth}
  \centering
  \includegraphics[width=.98\linewidth]{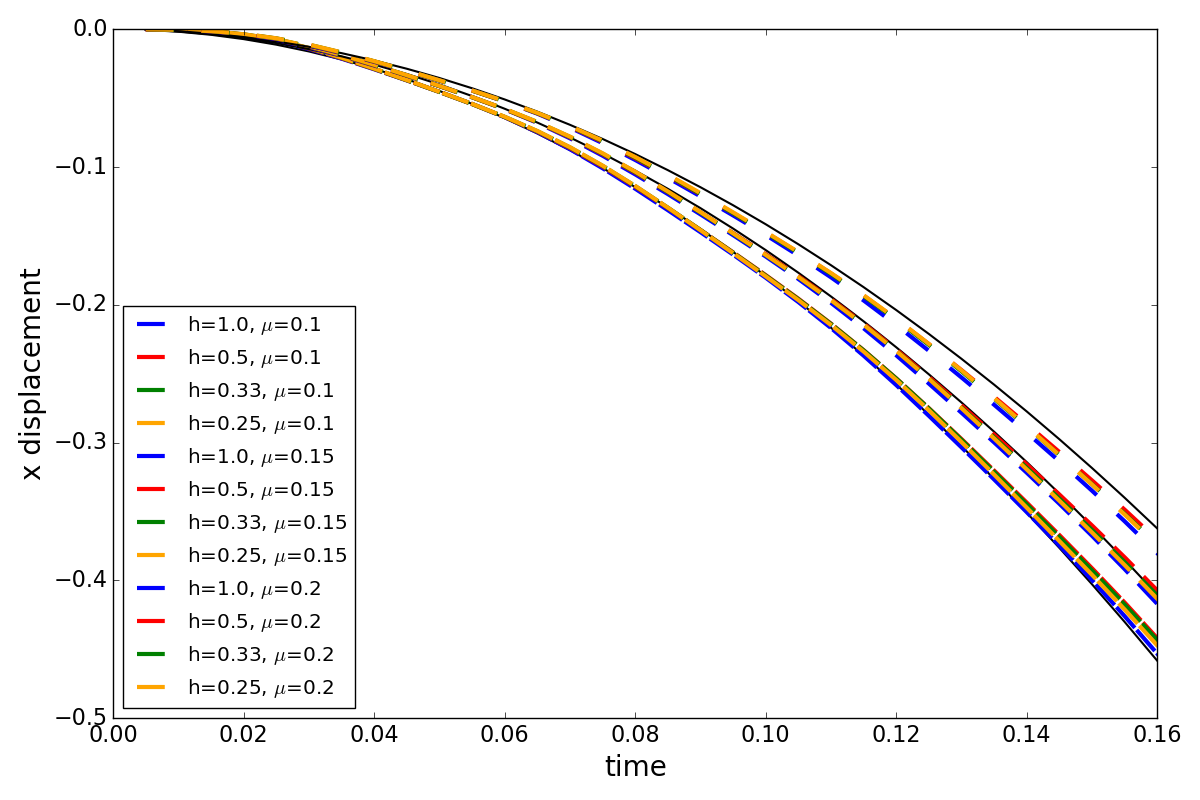}  
  \caption{PIC horizontal sliding compared to analytic solution (in black).}
\end{subfigure} \\
\begin{subfigure}{.6\textwidth}
  \centering
  \includegraphics[width=.98\linewidth]{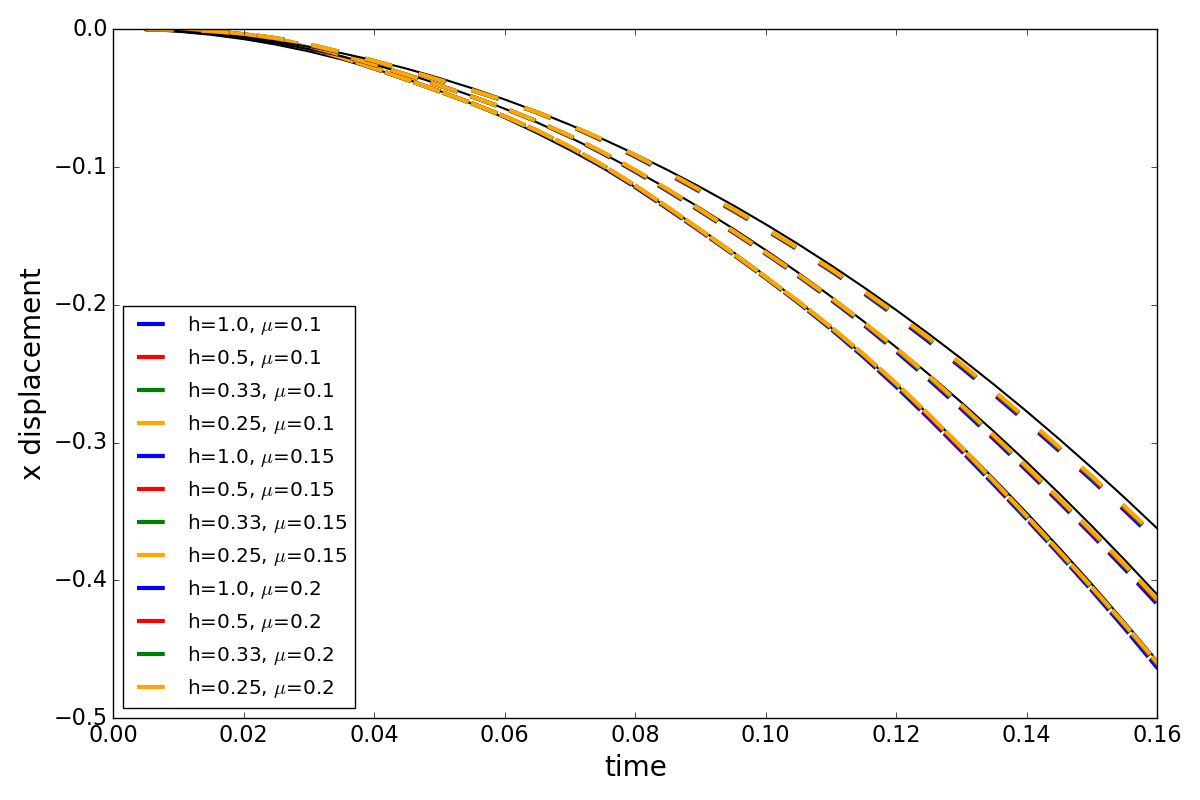}  
  \caption{APIC horizontal sliding compared to analytic solution (in black).}
\end{subfigure}
\caption{Displacement in x for a block sliding down a ramp with varying friction coefficient.}
\label{fig:SlidingBlockDisp}
\end{figure}

A frictional sliding example demonstrates the method for the case of sliding with frictional contact.  The geometry consists of a 1x1x1 unit cubic block sliding down a ramp which is tilted at a slope of 30$^\circ$, as shown in Figure~\ref{fig:SlidingBlockSnap}.   The ramp is 4 units long, 1.5 wide, and 0.25 thick.  The bottom face of the ramp is fully fixed.  The density, Young's modulus, and Poisson's ratio for both ramp and block are $1e6$, $1e10$ and $0.3$ respectively.  A gravity load of $g=-100$ is applied, and the time step is chosen to be $\Delta t = 2.5 \times 10^{-5} h$, for approximate mesh size $h=\{0.25, 0.33, 0.5, 1.0\}$.  The MPM grid size is taken to be $0.5h$, and the augury time is taken as $\tau = \Delta t$.  Results showing the displacement history of the block for varying mesh sizes and friction coefficients are shown in Figure~\ref{fig:SlidingBlockDisp}.  The analytic solution is also shown in black.  The results using both an PIC and APIC transfer are shown.  Clearly the use of APIC helps improve accuracy for these cases, and there appears to be improvement toward the exact solution as the mesh and background grid are refined.  

\section{Conclusions} \label{sec:Conclusions}

The proposed APIC contact algorithm has two clear advantages over established contact algorithms based on PIC methods: contact enforcement at near zero gap, and exact conservation of both linear and angular momentum.  Furthermore, the proposed algorithm has two advantages over established contact algorithms for explicit dynamic finite element simulations.  PIC-based contact algorithms require no object-to-object search, as all information is simply exchanged through a common background grid.  PIC-based algorithms have linear computational complexity and are naturally both shared memory and distributed memory scalable.  These two performance advantages results in a more efficient algorithm for enforcing contact than is traditionally possible.  While the novel ideas presented here show significant promise, opportunities for improvement remain.  Ongoing work includes: continuing to analyze the theoretical properties of the algorithm, and minimizing energy dissipation by generalizing the algorithm to work effectively even when the background grid is small relative to foreground mesh.

\section*{Acknowledgments}

Sandia National Laboratories is a multi-mission laboratory managed and operated by National Technology and Engineering Solutions of Sandia, LLC, a wholly owned subsidiary of Honeywell International Inc., for the U.S. Department of Energy's National Nuclear Security Administration under contract DE-NA0003525.  This paper describes objective technical results and analysis. Any subjective views or opinions that might be expressed in the paper do not necessarily represent the views of the U.S. Department of Energy or the United States Government.

\section*{Conflict of interest}
On behalf of all authors, the corresponding author states that there is no conflict of interest.

\begin{appendices}

\section{Linear completeness}
  
\label{subsec:linear_fields}

We prove one of the properties claimed in Section~\ref{sec:Formulation}: that affine velocity fields are preserved
under grid transfer operation $\vect{H}^{APIC}$.  First we have to establish two standard properties of our background grid approximations, namely partition of unity:
\eqlabel{eq:pou}{
  \sum_i N_i(\vect{x}) = 1,
}
and linear completeness:
\eqlabel{eq:linearCompleteness}{
  \sum_i N_i(\vect{x}) \vect{x}_i = \vect{x}.
}
The specific form of quadratic B-Splines which we use are described in~\ref{subsec:spline}.
Suppose each particle velocity starts with an affine
velocity field $\tilde{\vect{v}}(\vect{x}) = \vect{v}_0 + \vect{L} \vect{x}$, where $v_0$ is the velocity at $\vect{x}=\vect{0}$ and $\vect{L}$ is a constant velocity gradient tensor,
so
\eq{
  {\vect{v}}_p = \vect{v}_0 + \vect{L} \vect{x}_p, }
and
\eq{ {\vect{{B}}}_p = \vect{L}  \vect{D}_p. }
The particle-to-grid transfer results in grid velocities
\eq{
  m_i \vect{v}_i &= \sum_p N_i( \vect{x}_p ) m_p \left[ \vect{v}_0 + \vect{L} \vect{x}_p + \vect{L} \left( \vect{x}_i - \vect{x}_p  \right) \right] \nonumber \\
  &= \sum_p N_i( \vect{x}_p ) m_p \left[ \vect{v}_0 + \vect{L} \vect{x}_i  \right] \nonumber \\
  &= \sum_p N_i( \vect{x}_p ) m_p \tilde{\vect{v}}( \vect{x}_i ) \nonumber \\
  &= m_i \tilde{\vect{v}}( \vect{x}_i ).
}
The subsequent grid-to-particle transfer then gives
\eq{
  \vect{v}_p &= \sum_i N_i(\vect{x}_p) \vect{v}_i \nonumber \\
  &= \sum_i N_i(\vect{x}_p) \tilde{\vect{v}}( \vect{x}_i ) \nonumber \\
  &= \sum_i N_i(\vect{x}_p) \left( \vect{v}_0 + \vect{L} \vect{x}_i \right) \nonumber \\
  &= \vect{v}_0 + \vect{L} \vect{x}_p \nonumber \\
  &= \tilde{\vect{v}}(\vect{x}_p),
}
where we use~\eqref{eq:pou} and~\eqref{eq:linearCompleteness}.  Finally we confirm that the scaled velocity gradient term is unchanged by the grid transfers:
\eq{
  \vect{B}_p &= \sum_i N_i(\vect{x}_p) \tilde{\vect{v}}(\vect{x}_i) \otimes \left( \vect{x}_i - \vect{x}_p \right) \nonumber \\
  &= \sum_i N_i(\vect{x}_p) \left( \vect{v}_0 + \vect{L} \vect{x}_i \right) \otimes \left( \vect{x}_i - \vect{x}_p \right) \nonumber \\
  &= \vect{v}_0  \otimes \sum_i N_i(\vect{x}_p)  \left( \vect{x}_i - \vect{x}_p \right)  + 
  \sum_i N_i(\vect{x}_p) \vect{L} \vect{x}_i  \otimes \left( \vect{x}_i - \vect{x}_p \right) \nonumber \\
  &= \vect{v}_0  \otimes \left( \vect{x}_p - \vect{x}_p \right)  + 
  \vect{L} \sum_i N_i(\vect{x}_p) \vect{x}_i  \otimes \left( \vect{x}_i - \vect{x}_p \right) \nonumber \\
  &=  \vect{L}  \sum_i N_i(\vect{x}_p)\vect{x}_i  \otimes \left( \vect{x}_i - \vect{x}_p \right) \nonumber \\
  &=  \vect{L}  \left( \sum_i N_i(\vect{x}_p)\vect{x}_i - \vect{x}_p \right)  \otimes \left( \vect{x}_i - \vect{x}_p \right) \nonumber \\
  &=  \vect{L}  \vect{D}_p,
}
where we use 
\eq{
  \sum_i N_i(\vect{x}_p)  \vect{x}_p   \otimes \left( \vect{x}_i - \vect{x}_p \right) = \vect{x}_p \otimes \sum_i N_i(\vect{x}_p) \left( \vect{x}_i - \vect{x}_p \right)
  = \vect{x}_p \otimes \left( \vect{x}_p - \vect{x}_p \right) = 0.
}
This proves that affine velocity fields are preserved by APIC grid transfers.

\section{Quadratic B-Spline Basis}
\label{subsec:spline}

For the background grid approximation space, quadratic B-splines are employed.  We adjust the standard quadratic B-splines slightly to ensure both partition of unity and linear completeness~\cite{Zhang:2017aa}, as is required for the proof in~\ref{subsec:linear_fields}.  Only the 1D shape functions are defined here.  The full 3D approximation space is constructed from the standard outer product of 1D spline spaces.  For a point $x$ in a 1D grid with grid size $h$, we compute the grid point location just to the left as
\eq{
x_i = h \; \text{floor}\left( \frac{x}{h} \right).
}
Using \eq{ \xi = \frac{x}{h} \text{  and  } i = \frac{x_i}{h},} the shape functions which cover $x$ are
\eq{
N_{i-1}(\xi) &= \frac{1}{4}a_i(\xi) \nonumber \\
N_i(\xi) &= \frac{1}{8}\left( a_i(\xi) + b_i(\xi) \right) \nonumber \\
N_{i+1}(\xi) &= \frac{1}{8}\left( b_i(\xi) + c_i(\xi) \right) \nonumber \\
N_{i+2}(\xi) &= \frac{1}{4} c_i(\xi),
}
with 
\eq{
a_i(\xi) &= (i + 1 - \xi)^2, \nonumber \\
b_i(\xi) &= (-i + 1 + \xi)  (i + 1 - \xi) + (i + 2 - \xi) (\xi - i), \nonumber \\
c_i(\xi) &= (\xi - i )^2.
}
For this background grid, the value for the tensor $\vect{D}_p$ associated with computing the scaled velocity gradient tensor from Section~\ref{subsec:apic_formulation} is constant over the entire grid, and has value $\vect{D}_p = \frac{1}{2} h^2 \vect{I}$.

\end{appendices}

\comment{

\subsection{Conservation of linear momentum}
We proceed to prove conservation of linear and angular momentum.  These proofs follow those in~\cite{JIANG2017137}, and
are included here solely for completeness.  Staring with linear momentum we have
\eq{
  \vect{p}^G &= \sum_i m_i \vect{v}_i \nonumber \\
  &= \sum_i \sum_p N_i(\vect{x}_p) m_p \left( \vect{v}_p + \vect{B}_p \vect{D}_p^{-1} \left( \vect{x}_i - \vect{x}_p \right) \right)  \nonumber \\
  &= \sum_i \sum_p N_i(\vect{x}_p) m_p  \vect{v}_p + \sum_i \sum_p N_i(\vect{x}_p) m_p \vect{B}_p \vect{D}_p^{-1} \left( \vect{x}_i - \vect{x}_p \right)  \nonumber \\
  &=  \sum_p  m_p \vect{v}_p \left( \sum_i   N_i(\vect{x}_p) \right)  + \sum_p  m_p \vect{B}_p \vect{D}_p^{-1} \left( \sum_i  N_i(\vect{x}_p) \left( \vect{x}_i - \vect{x}_p \right) \right) \nonumber \\
  &= \sum_p m_p \vect{v}_p + \vect{0} \nonumber \\
  &= \vect{p}^P.
}
where $\vect{P}^G$ and $\vect{P}^P$ are grid and particle momentum respectively.  We now need to confirm that the grid-to-particle transfer similarly conserves linear momentum:
\eq{
  \vect{p}^P &= \sum_p m_p \vect{v}_p\nonumber \\
  &= \sum_p m_p \left( \sum_i N_i(\vect{x}_p) \vect{v}_i \right) \nonumber \\
  &= \sum_i \left( \sum_p N_i(\vect{x}_p) m_p \right) \vect{v}_i  \nonumber \\
  &= \sum_i m_i \vect{v}_i \nonumber \\
  &= \vect{p}^G.
}

}

\bibliographystyle{spmpsci}


\end{document}